\documentclass[acmsmall]{acmart}
\AtBeginDocument{%
}

\setcopyright{rightsretained}
\acmJournal{PACMMOD}
\acmYear{2023} \acmVolume{1} \acmNumber{2} \acmArticle{198} \acmMonth{6} \acmPrice{}\acmDOI{10.1145/3589778}

\makeatletter
\gdef\@copyrightpermission{
 \begin{minipage}{0.2\columnwidth}
  \href{https://creativecommons.org/licenses/by/4.0/}{\includegraphics[width=0.90\textwidth]{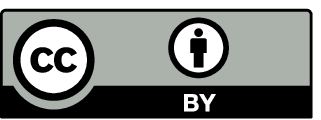}}
 \end{minipage}\hfill
 \begin{minipage}{0.8\columnwidth}
  \href{https://creativecommons.org/licenses/by/4.0/}{This work is licensed under a Creative Commons Attribution International 4.0 License.}
 \end{minipage}
 \vspace{5pt}
}
\makeatother

\usepackage{mathbbol}
\usepackage{tikz}
\usetikzlibrary{arrows,arrows.meta,calc}
\usetikzlibrary{shadows}
\usepackage{enumerate}
\usepackage{xspace}
\usepackage{hyperref}
\usepackage{xurl}
\usetikzlibrary{shapes.multipart}
\usetikzlibrary{positioning}
\usetikzlibrary{shadows}
\usetikzlibrary{shapes.multipart}
\usepackage{xcolor-solarized}
\usepackage{booktabs} %for nicer looking tables
\usepackage{enumitem} %for making itemize more compact
\usepackage{multirow}
\usepackage{listings}
\usepackage{appendix}
\lstdefinelanguage{pgkeysLang}
{
  morekeywords={
    NODE,
    EDGE,
    TYPE,
    IMPORTS,
    OPTIONAL,
    DATE,
    INT,
    STRING,
    BOOL,
    ARRAY,
    ENUM,
    DOUBLE,
    INT32,
    DATETIME,
    STRICT,
    LOOSE,
    OPEN,
    CLOSED,
    ABSTRACT,
    CREATE,
    GRAPH,
    FOR,
    WITHIN,
    EXCLUSIVE,
    MANDATORY,
    SINGLETON,
    FLOAT,
    IDENTIFIER,
    MATCH,
    WHERE,
    NOT,
    OR,
    AND,
    EXISTS,
    RETURN,
    IN,
    IS,
    NULL,
    hasKey,
    FunctionalProperty,
    context,
    test,
    unique,
    selector,
    field,
    allInstances,
    isUnique,
    Tuple,
    class,
    extent,
    relationship,
    inverse,
    attribute,
    COUNT,
    OF
  },
  sensitive=false, % keywords are not case-sensitive
  morecomment=[l]{//}, % l is for line comment
  morecomment=[s]{/*}{*/}, % s is for start and end delimiter
  morestring=[b]" % defines that strings are enclosed in double quotes
}

\usepackage{color}
\definecolor{eclipseBlue}{RGB}{42,0.0,255}
\definecolor{eclipseGreen}{RGB}{63,127,95}
\definecolor{eclipsePurple}{RGB}{127,0,85}
 
\newcommand{\inlinecode}[1]{{\small {\ttfamily #1}}}

\usepackage{calc}
\usepackage{xcolor-solarized}

\newcommand{\noop}[1]{}

\makeatletter
\newcommand{\pto}{}% just for safety
\DeclareRobustCommand{\pto}{\mathrel{\mathpalette\p@to@gets\to}}
\newcommand{\p@to@gets}[2]{%
  \ooalign{\hidewidth$\m@th#1\mapstochar\mkern5mu$\hidewidth\cr$\m@th#1\to$\cr}%
}
\makeatother

\usepackage{array} % for extended column definitions    
\usepackage{arydshln} % for dashed / dotted lines in tables

\usepackage{adjustbox}  % allows to shrink a table to fit, if necessary

\newcommand{\OMIT}[1]{}
\newcommand{\dom}{\ensuremath{\mathrm{dom}}}

\newcommand{\BB}{\ensuremath{\mathcal{B}}}

\newcommand{\KK}{\ensuremath{\mathcal{K}}}
\newcommand{\LL}{\ensuremath{\mathcal{L}}}

\newcommand{\RR}{\ensuremath{\mathcal{R}}}
\newcommand{\TT}{\ensuremath{\mathcal{T}}}
\newcommand{\VV}{\ensuremath{\mathcal{V}}}

\newcommand{\pgschema}{\textsc{PG-Schema}\xspace}
\newcommand{\pgtypes}{\textsc{PG-Types}\xspace}

\newcommand{\pgkeys}{\textsc{PG-Keys}\xspace}
\newcommand{\pgkey}{\textsc{PG-Key}\xspace}

\newcommand{\nsem}[1]{\Lparen #1 \Rparen}
\newcommand{\esem}[1]{\llbracket #1 \rrbracket}

\newcommand{\basetype}{\mathbb{b}}
\newcommand{\extbasetype}{\mathbb{B}}

\newcommand{\comma}{{\normalsize\textrm{\:,}}}
\newcommand{\period}{{\normalsize\textrm{\:.}}}

\newcommand{\omitforcameraready}[1]{}

\newcommand{\req}[1]{\textbf{R#1}}
\newcommand{\types}{\ensuremath{\mathsf{Types}}}

\newcommand{\mcomment}[2]{{\color{blue}\textbf{(#1)}}\footnote{\textbf{#1:} #2}}
\newcommand{\angela}[1]{\mcomment{Angela}{#1}}

\newcommand{\filip}[1]{\mcomment{Filip}{#1}}

\newcommand{\wim}[1]{\mcomment{Wim}{#1}}

\newcommand{\slawek}[1]{\mcomment{Slawek}{#1}}

\newcommand{\lacks}{-} % mark in table for lack of support
\newcommand{\qual}[1]{[#1]} % qualification of mark as partial or needing explanation
\newcommand{\has}{\ensuremath{\checkmark}} % mark in table for lack of support
\newcommand{\unkn}{?} % mark in table to indicate it cannot be determined if feature is supported
\begin{document}

%%
%% The "title" command has an optional parameter,
%% allowing the author to define a "short title" to be used in page headers.
\title{\pgschema: Schemas for Property Graphs}

%%
%% The "author" command and its associated commands are used to define
%% the authors and their affiliations.
%% Of note is the shared affiliation of the first two authors, and the
%% "authornote" and "authornotemark" commands
%% used to denote shared contribution to the research.

\author{Renzo Angles}
\affiliation{
\institution{Faculty of Engineering, Universidad de Talca}
\city{Curicó}
\country{Chile}
}
\orcid{0000-0002-6740-9711}
\email{rangles@utalca.cl}

\author{Angela Bonifati}
%\affiliation{
%\institution{Lyon 1 University \& Liris CNRS}
%\country{France}
%}
\affiliation{
\institution{Lyon 1 University \&  Liris CNRS}
\city{Villeurbanne}
\country{France}
}
\orcid{0000-0002-9582-869X}
\email{angela.bonifati@univ-lyon1.fr}

\author{Stefania Dumbrava}
\affiliation{
\institution{ENSIIE \& SAMOVAR - Institut Polytechnique de Paris}
\city{Paris}
\country{France}
}
\orcid{0000-0002-6664-0620}
\email{stefania.dumbrava@ensiie.fr}

\author{George Fletcher}
\affiliation{
\institution{Eindhoven University of Technology}
\city{Eindhoven}
\country{Netherlands}
}
\orcid{0000-0003-2111-6769}
\email{g.h.l.fletcher@tue.nl}

\author{Alastair Green}
\affiliation{
\institution{LDBC}
\city{London}
\country{UK}
}
\orcid{0000-0002-3166-6708}
\email{alastair@acm.org}

\author{Jan Hidders}
\affiliation{
\institution{Birkbeck, University of London}
\city{London}
\country{UK}
}
\orcid{0000-0002-8865-4329}
\email{j.hidders@bbk.ac.uk}

\author{Bei Li}
\affiliation{
\institution{Google}
\city{Mountain View}
\country{USA}
}
\orcid{0009-0002-8102-978X}
\email{bei@google.com}

\author{Leonid Libkin}
%\authornote{Also affiliated with ENS, PSL  University, Paris, France.}
\affiliation{
\institution{University of Edinburgh}
\city{Edinburgh}
\country{UK}
}
\affiliation{
\institution{RelationalAI \& ENS, PSL University}
\city{Paris}
\country{France}
}
\orcid{0000-0002-6698-2735}
\email{l@libk.in}

\author{Victor Marsault}
\affiliation{
\institution{LIGM, Université Gustave Eiffel, CNRS}
\city{Champs-sur-Marne}
\country{France}
}
\orcid{0000-0002-2325-6004}
\email{victor.marsault@univ-eiffel.fr}

\author{Wim Martens}
\affiliation{
\institution{University of Bayreuth}
\city{Bayreuth}
\country{Germany}
}
\orcid{0000-0001-9480-3522}
\email{wim.martens@uni-bayreuth.de}

\author{Filip Murlak}
\affiliation{
\institution{University of Warsaw}
\city{Warsaw}
\country{Poland}
}
\orcid{0000-0003-0989-3717}
\email{f.murlak@uw.edu.pl}

\author{Stefan Plantikow}
\affiliation{
\institution{Neo4j}
\city{Berlin}
\country{Germany}
}
\orcid{0009-0006-2910-1050}
\email{stefan.plantikow@neo4j.com}

\author{Ognjen Savkovi\'c}
\affiliation{
\institution{Free University of Bozen-Bolzano}
\city{Bolzano}
\country{Italy}}
\orcid{0000-0002-9141-3008}
\email{ognjen.savkovic@unibz.it}

\author{Michael Schmidt}
\affiliation{
\institution{Amazon Web Services}
\city{Seattle}
\country{USA}}
\orcid{0009-0002-3292-0349}
\email{schmdtm@amazon.com}

\author{Juan Sequeda}
\affiliation{
\institution{data.world}
\city{Austin}
\country{USA}
}
\orcid{0000-0003-3112-9299}
\email{juan@data.world}

\author{S\l{}awek Staworko}
\affiliation{
\institution{RelationalAI}
\city{Berkeley}
\country{USA}
}
\affiliation{
\institution{Univ. Lille, CNRS, UMR 9189 CRIStAL}
\city{F-59000 Lille}
\country{France}
}
\orcid{0000-0003-3684-3395}
\email{slawek.staworko@relational.ai}

\author{Dominik Tomaszuk}
\affiliation{
\institution{University of Bialystok}
\city{Bialystok}
\country{Poland}
}
\orcid{0000-0003-1806-067X}
\email{d.tomaszuk@uwb.edu.pl}

\author{Hannes Voigt}
\affiliation{
\institution{Neo4j}
\city{Leipzig}
\country{Germany}
}
\orcid{0000-0002-2148-9592}
\email{hannes.voigt@neo4j.com}

\author{Domagoj Vrgo\v{c}}
\affiliation{
\institution{University of Zagreb}
\city{Zagreb}
\country{Croatia}
}
\affiliation{
\institution{PUC Chile}
\city{Santiago de Chile}
\country{Chile}
}
\orcid{0000-0001-5854-2652}
\email{vrdomagoj@uc.cl}

\author{Mingxi Wu}
\affiliation{
\institution{TigerGraph}
\city{Redwood City}
\country{USA}
}
\orcid{0009-0009-2738-7018}
\email{mingxi.wu@tigergraph.com}

\author{Dušan Živković}
\affiliation{
\institution{Integral Data Solutions}
\city{London}
\country{UK}
}
\orcid{0009-0003-4658-3774}
\email{dusan.zivkovic@me.com}

%%
%% By default, the full list of authors will be used in the page
%% headers. Often, this list is too long, and will overlap
%% other information printed in the page headers. This command allows
%% the author to define a more concise list
%% of authors' names for this purpose.
\renewcommand{\shortauthors}{Renzo Angles et al.}
%% No italics and no comma
%% If needed use a foot or author note to identify equal contribution

%%
%% The abstract is a short summary of the work to be presented in the
%% article.
%%
%% The abstract is a short summary of the work to be presented in the
%% article.
\begin{abstract}
  %We propose a formalism for specifying property graph schemas. We keep it simple.
  Property graphs have reached a high level of maturity, witnessed by multiple robust graph database systems as well as the ongoing ISO standardization effort aiming at creating a new standard Graph Query Language (GQL). Yet, despite documented demand, schema support is limited both in existing systems and in the first version of the GQL Standard. It is anticipated that the second version of the GQL Standard will include a rich DDL. Aiming to inspire the development of GQL and enhance the capabilities of graph database systems, we propose \pgschema, a simple yet powerful formalism for specifying property graph schemas. It features \pgtypes with flexible type definitions supporting multi-inheritance, as well as expressive constraints  based on the recently proposed  \pgkeys formalism. We provide the formal syntax and semantics of \pgschema, which meet principled design requirements grounded in contemporary property graph management scenarios, and offer a detailed comparison of its features with those of existing schema languages and graph database systems. 
\end{abstract}

%%% Local Variables:
%%% mode: latex
%%% TeX-master: "sample-authordraft"
%%% End:

%%
%% The code below is generated by the tool at http://dl.acm.org/ccs.cfm.
%% Please copy and paste the code instead of the example below.
%%
\begin{CCSXML}
<ccs2012>
<concept>
<concept_id>10002951.10002952.10003190.10003206</concept_id>
<concept_desc>Information systems~Integrity checking</concept_desc>
<concept_significance>500</concept_significance>
</concept>
<concept>
<concept_id>10003752.10010070.10010111.10010112</concept_id>
<concept_desc>Theory of computation~Data modeling</concept_desc>
<concept_significance>500</concept_significance>
</concept>
<concept>
<concept_id>10003752.10010070.10010111.10010114</concept_id>
<concept_desc>Theory of computation~Database constraints theory</concept_desc>
<concept_significance>500</concept_significance>
</concept>
</ccs2012>
\end{CCSXML}

\ccsdesc[500]{Information systems~Integrity checking}
\ccsdesc[500]{Theory of computation~Data modeling}
\ccsdesc[500]{Theory of computation~Database constraints theory}

%%
%% Keywords. The author(s) should pick words that accurately describe
%% the work being presented. Separate the keywords with commas.
\keywords{property graphs; schemas; graph databases}

\received{November 2022}
\received[revised]{February 2023}
\received[accepted]{March 2023}

%%
%% This command processes the author and affiliation and title
%% information and builds the first part of the formatted document.
\maketitle

\section{Introduction}
\label{sec:introduction}

%P1: PG ARE MAINSTREAM
Property graphs have come of age. 
The property graph data model is widely used in social and transportation networks, biological networks, finance, cyber security, logistics, and planning domains to represent interconnected multi-labeled data enhanced with properties given by key/value pairs \cite{SakrBVIAAAABBDV21}.
Its maturity is reflected in the ongoing  efforts by ISO (International Organization for Standardization)\footnote{ISO's Working Group for Database Languages is known as ISO/IEC JTC1 SC32 WG3.} to create a standard Graph Query Language GQL, which is expected to appear in 2024 \cite{ISO39075,DeutschFGHLLLMM22}. 
%%JUAN NOTE: BRINGING UP SQL/PGQ WOULD MAKE THE CONTEXT COMPLETE, BUT IT IS NOT NEEDED FOR THIS PAPER BECAUSE OUR FOCUS IS GQL, SO BEST TO IGNORE IT. IF I WOULD BE AN ASSHOLE REVIEWER, I WOULD BRING UP THE NEED TO COMPARE PG-SCHEMA WITH THE SCHEMAS IN SQL/PGQ. LET'S AVOID THIS
%which currently develops two graph data standards: a standalone  and SQL's extension for querying property graphs SQL/PGQ. These standards, to appear in 2023 and 2024, are based on a common data model and a common graph pattern matching language; see 

%P2: PG DON'T HAVE SCHEMAS, BUT THEY ARE NEEDED
Despite the maturity of commercial and open-source property graph databases, their schema support is limited.
Schemas are a fundamental building block for many data systems. They provide structure to data in a formal language, and are used in different scenarios.  
\OMIT{
\begin{itemize}
\item In the \emph{schema-first} scenario, dominating in production settings of stable systems, the schema is provided during the setup and plays a \emph{prescriptive role}, limiting data modifications.
\item In the \emph{flexible schema} scenario, convenient for rapid application development and data integration, schema information comes together with data and plays a \emph{descriptive role}, telling users and systems what to expect in the data. 
\item In the \emph{partial schema} scenario, applicable at advanced development stages, the user wants to enforce a prescriptive schema over parts of the data corresponding to the stable parts of the data model, and maintain a descriptive schema depicting the whole data, including its quickly evolving parts. 
\end{itemize}
}
% end omit itemized 
%
In the \emph{schema-first} scenario, dominating in production settings of stable systems, the schema is provided during the setup and plays a \emph{prescriptive role}, limiting data modifications.
In the \emph{flexible schema} scenario, suitable for rapid application development and data integration, schema information comes together with data and plays a \emph{descriptive role}, telling users and systems what to expect in the data. 
In the \emph{partial schema} scenario, applicable at advanced development stages, the user wants to enforce a prescriptive schema over stable parts of the data
and maintain a descriptive schema depicting the whole data including its
evolving parts. 

A recent survey of graph processing users \cite{SahuMSLO20} revealed that schema compliance is a highly desirable feature that is lacking in property graph database systems. 
Our inspection of eleven property graph engines reveals a fragmented landscape, where no system offers comprehensive support for schemas, which should allow the user to impose structure on nodes, edges, and properties of the underlying graph instances, as well as enforce constraints. 
This calls for a unified property graph schema language.  

%P3: OUR GOAL: PROVIDE A PROPOSAL FOR ISO/GQL TO SPEED UP DEVELOPMENT
Our goal is to support the endeavours surrounding the GQL standard and accelerate the development of a future standardized property graph schema language by presenting a concrete proposal.
Our proposal, called \pgschema, consolidates and extends discussions arising out of the Property Graph Schema Working Group of the Linked Data Benchmark Council \cite{LDBC:OAEP:OAEP-2023-04}. 
This model of providing recommendations to standards committees has proven successful, as evidenced by G-CORE \cite{gcore} and \pgkeys \cite{PGkeys} influencing GQL\footnote{Seven authors of this paper are also members of %ISO's Working Group for Database Languages.
ISO/IEC JTC1 SC32 WG3.}. 

\begin{figure}[t]
\centering
    %\resizebox{\linewidth}{!}{
    \footnotesize
    \begin{tikzpicture}[
      Node/.style={
        rounded corners,
        rectangle split,
        rectangle split parts=2,
        rectangle split part fill={green!80!black!15!white,solarized-base3!50!white},
        draw=green!20!black,
        semithick
      },
      ISA/.style={
        thick,-{Triangle[open]},
      },
      Edge/.style={
        thick,-stealth',
      },
      EdgeProperties/.style={
        sloped,
        below,
        rounded corners,
        fill=solarized-base3!50!white,
      }
      ]

      \path[use as bounding box] (-4,-0.5) rectangle (-16,-4.5);

      \node[Node] (person) at (-6, -1.5) {
        \texttt{Person}
        \nodepart{two}
        \begin{tabular}{l}
        \textbf{name} {STRING}
        \end{tabular}
      };

      \node[Node] (customer) at (-10, -1.5) {
        \texttt{Customer}
        \nodepart{two}
        \begin{tabular}{l}
        \textbf{id} {INT32}
        \end{tabular}
      };
      
      \node[Node] (credit_card) at (-6,-4)  {
        \texttt{CreditCard}
        \nodepart{two}
        \begin{tabular}{l}
        \textbf{num} {STRING}
        \end{tabular}
      };

      \node[Node] (account) at (-14, -1.5) {
        \texttt{Account}
        \nodepart{two}
        \begin{tabular}{l}
        \textbf{name} {STRING}
        \end{tabular}
      };

      \node[Node] (transaction) at (-14,-4) {
        \texttt{Transaction}
        \nodepart{two}
        \begin{tabular}{l}
        \textbf{num} {STRING}
        \end{tabular}
      };

      \draw[] (customer) edge[ISA] 
      (person);

      \draw[] (customer) edge[Edge] 
      node[above] {\texttt{owns}} 
      (account);

      \draw[] (customer) edge[Edge,bend right=15] 
      node[above right] {\texttt{uses}} (credit_card);

      \draw[] (transaction) edge[Edge] 
      node[above] {\texttt{charges}}
      node[below,EdgeProperties] {\textbf{amount} {DOUBLE}}
      (credit_card);

      \draw[] (transaction) edge[Edge,bend left] 
      node[left] {\texttt{deposits}}
      (account);
     
      \draw[] (transaction) edge[Edge,bend right] 
      node[right] {\texttt{withdraws}}
      (account);

    \end{tikzpicture}
  %}

%%% Local Variables:
%%% mode: latex
%%% TeX-master: "sample-authordraft"
%%% End:
\caption{A  diagram of a fraud graph schema. }
\label{fig-uc2-diagram}
\end{figure}
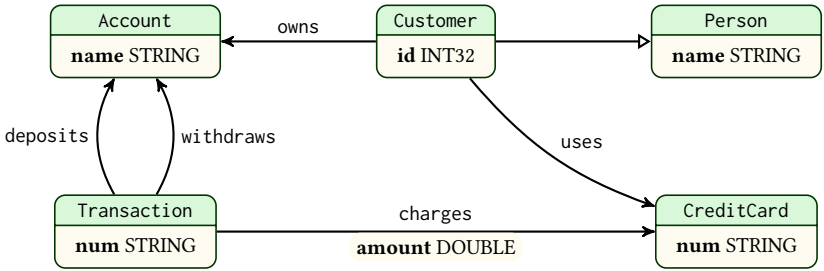

\begin{figure}[t!]
\begin{lstlisting}
                CREATE GRAPH TYPE   fraudGraphType STRICT {
                  (personType: Person {name STRING}),
                  (customerType: personType & Customer {id INT32}),
                  (creditCardType: CreditCard {num STRING}),
                  (transactionType: Transaction {num STRING}),
                  (accountType: Account {id INT32}),
                  (:customerType)
                    -[ownsType: owns]->
                  (:accountType),
                  (:customerType)
                    -[usesType: uses]->
                  (:creditCardType),
                  (:transactionType)
                    -[chargesType: charges {amount DOUBLE}]->
                  (:creditCardType),
                  (:transactionType)
                    -[activityType: deposits|withdraws]->
                  (:accountType)
                }
\end{lstlisting}
%\vspace{-10pt}
\caption{\pgschema{} of a fraud graph schema.}
\label{fig-uc2}
\end{figure}

%%%%% FULL PG-SCHEMA just in case. 
\OMIT{
  
CREATE GRAPH TYPE fraudGraphType STRICT {

    (personType: Person {name STRING}),
    (customerType: personType & Customer {id INT32}),
    (transactionType: Transaction {num STRING}),
    (accountType: Account {id INT32}),
    ...
    (:transactionType)-[activityType: deposit|withdraw]->(:accountType)
}
}

%P4: MOTIVATING EXAMPLE, HIGHLIGHTING KEY SCHEMA FEATURES
To illustrate the key features required of schemas for property graphs, we now provide a concrete example of a schema in fraud detection, a common application of graph databases \cite{PractGraph},
and show how the schema can be used in interactive graph exploration, which in itself is a common functionality provided by graph databases~\cite{TSDB, GraphNB, NeoB, TigGraph, Grafo}.  
The diagram in Figure~\ref{fig-uc2-diagram} represents a schema describing a fraud graph.
%while Figure~\ref{fig-uc2} describes it in the syntax of \pgschema. 
This schema is used by Andrea, who works as a financial compliance officer and utilizes an interactive graph explorer to investigate fraud. 
The following  user session highlights how schema information can enhance Andrea’s user experience. (The scenario is highly simplified and does not reflect the actual complexity of such tasks and applicable techniques.)
\begin{enumerate}
\item Andrea opens the graph explorer and connects to the fraud graph, one of the resources in the data catalog.  While establishing the connection, the application bootstraps by loading schema information. \item Andrea first seeks to identify pairs of suspicious customers. Being aware of the schema, the graph explorer leverages the node type definitions to construct a start page that proposes search for any of the entities available in the domain: \inlinecode{Customer}, \inlinecode{Account}, etc.  
Andrea proceeds with \inlinecode{Customer}.
\item Based on the schema information, the graph explorer dynamically constructs a 
\inlinecode{Customer} search form. It contains separate search fields for the known properties of \inlinecode{Customer} nodes: in our example \inlinecode{id} and \inlinecode{name} (inherited from \inlinecode{Person}).
Based on the data type constraints in the schema, the \inlinecode{id} field accepts only integers as input. Andrea uses the \inlinecode{name} field to search for 
customers of interest and obtains a visual representation of the customer nodes in response.
\item 
To inspect potential fraudulent behavior, Andrea needs to understand the connections between customers. 
Exploiting the schema, the graph explorer leverages the edge type constraints to enumerate and propose specific types of connections. For instance,  knowing that \inlinecode{Customer}s use \inlinecode{CreditCard}s and \inlinecode{Transaction}s charge \inlinecode{CreditCard}s, it identifies 
structural connection patterns, written here in Cypher/GQL style,

\begin{lstlisting}
  (x:Customer)
   -[:uses]->(:CreditCard)<-[:uses]-
  (y:Customer),
\end{lstlisting}
\begin{lstlisting}
  (x:Customer)-[:uses]->(:CreditCard)
    <-[:charges]-(t:Transaction)-[:charges]->
  (:CreditCard)<-[:uses]-(y:Customer)
\end{lstlisting}

and lets Andrea choose the patterns of interest.
\item Andrea selects the first pattern, which aims to identify shared credit cards that were used by two customers. Based on the schema knowledge, the application constructs and executes an efficient query that quickly identifies all shared credit cards usages between customers.
\item  The graph explorer visualizes the results of the connection search. Upon further investigation, Andrea confirms suspicious cross-customer usage of the same credit card and classifies the cases as fraudulent behavior.
\end{enumerate}
As illustrated by this sample session, the graph explorer would not have been able to effectively guide Andrea through the exploration without concrete schema information. The suggestion of property-specific search restrictions is made possible by {\em content types}. The schema-assisted query formulation leverages \emph{node} and \emph{edge types}. 
Going beyond our example, it is easy to see how other type and constraint information may help: for instance, key constraints could indicate preferred search fields; more general participation constraints may improve the schema-assisted query formulation process; orthogonal tooling for schema generation might give graph explorers a standardized path to approximate schema information in case it has not been provided by the graph database authors. These considerations lead to a number of {\em design requirements} (see Section \ref{sec:requirements}), upon which we base our proposal. 
The design requirements reflect the consensus of all authors, bringing to bear the theory and contemporary practice of graph schemas.

%\leonid{let's see if the space constraints eliminate the "due to space constraints" sentence}
%Due to space constraints, we discuss a second compelling application on schema-guided data catalogs, further corroborating the need of \pgschema{}, in an external technical report \footnote{\url{TODO - URL Appendix}}. 

% P5: OUR PROPOSAL
Our proposed \pgschema (\emph{Property Graph Schema Language}) comprises \pgtypes and \pgkeys.  \pgtypes specify possible combinations of labels and properties in nodes and edges of different types, as well as constraining the types of edges allowed between nodes of certain types, with the help of a rich inheritance mechanism and abstract types. 
%\pgtypes come with a parser \cite{grammar}. 
\pgkeys \cite{PGkeys} support diverse integrity  constraints, including keys and participation constraints. Via the mechanism of \emph{strict} and \emph{loose} schemas, and partial validation, \pgschema supports both the descriptive and prescriptive function of schemas.  No existing graph database system nor currently envisioned standard covers this arsenal of features; nor do they provide full schema validation support. Our \pgschema proposal provides both existing systems (reviewed in Section \ref{sec:discussion}) and forthcoming standards with these features, responding to users' demands.

% P6: PG-SCHEMA SAMPLE
We give a detailed description of \pgschema in Section~\ref{sec:proposal}. As a sample, consider the graph type definition in Figure~\ref{fig-uc2}, representing part of the schema diagram of Figure~\ref{fig-uc2-diagram}. 
The graph type \inlinecode{fraudGraphType} specifies node types (e.g., \inlinecode{personType}, \inlinecode{customerType}), and edge types (e.g., \inlinecode{activityType}) using the ASCII-art notation of Cypher \cite{cypher}, also adopted by GQL.
Properties are declared in curly braces \inlinecode{\{...\}}.
For example, the first statement defines a node type \inlinecode{personType} with label \inlinecode{Person} and a property with key \inlinecode{name} of type \inlinecode{STRING}; and the last statement defines an edge type \inlinecode{activityType}, whose label is either \inlinecode{deposits} or \inlinecode{withdraws}, which  connects nodes of types \inlinecode{transactionType} and \inlinecode{accountType}. 
 
\smallskip

\noindent{\bf Contributions.} 
We make the following contributions:
\begin{enumerate}
\item an analysis of the requirements for property graph schemas;
\item a proposal for a flexible, agile, usable, and expressive formalism called \pgschema{} that fulfills these requirements; 
\item full syntax and semantics of \pgschema, making the proposal easy to incorporate into both standards and systems; 
\item a parser for \pgschema \cite{grammar};
\item a detailed analysis of schemas in other structured and semi-structured data models and practical graph database systems, as well as their comparison with \pgschema.
\end{enumerate}
Our contributions impact the following audiences:
\begin{enumerate}[label=(\alph*)]
\item graph database standards committee members, who can build upon our recommendations for upcoming  features, 
\item graph database vendors, who can use our framework as a guideline to incorporate schemas in their systems, and
\item researchers, who can use a concrete model of schemas for property graphs as a basis for further investigation.
\end{enumerate}
\section{Design Requirements} 
\label{sec:requirements}
In this section, we elaborate on the design requirements for a suitable notion of a schema for property graphs. 
These requirements reflect a consensus reached in the course of a systematic multi-phase process informed by the scientific literature, the key use cases from industry, and the forthcoming GQL and SQL/PGQ standards \cite{ISO9075,ISO39075} (drafts of both standards are available to LDBC members).
%\cite{DeutschFGHLLLMM22}
%We first briefly discuss key characteristics of Property Graph Databases and then outline the relevant functions that schemas play in databases.

\subsection{Property Graphs and Database Schemas}
\label{subsec:PG-and-schemas}
Beyond the ubiquity of applications that focus on graph-structured data and graph analytics, the popularity of graph databases is also generally attributed to the following factors~\cite{Neo4j16,NeHo19,RoWeEi15,MSSJO20}. 
\begin{description}
\item \textsf{Agility.} Thanks to its proximity to conceptual data models, the property graph
  data model allows for an efficient translation from domain concepts to database
  items. This facilitates \textit{high responsiveness} i.e., the
  ability to quickly and reliably adapt to emerging organizational and domain needs, which is
  often achieved with iterative and incremental software development processes. 
\item \textsf{Flexibility.} Graph databases are aligned with an iterative and
  incremental development methodology because they do not require a rigid schema-based
  data governance mechanism, but rather favor test-driven development, which embraces
  the additive nature of graphs. The data does not need to be modeled in
  exhaustive detail in advance but rather \textit{new kinds of objects and
    relationships can emerge naturally} as new domain needs are addressed by
  evolving applications.
  % What makes graph databases aligned with iterative and incremental software
  % development is their schema-less nature. The data does not need to be modeled
  % in exhaustive detail in advance but rather \textit{new kinds of objects and
  %   relationships can emerge naturally} as new business needs are addressed by
  % evolving applications. A rigid schema-based data governance mechanisms are
  % forgone in favor of test-driven development, which embraces the additive
  % nature of graphs (new kinds of data can be added without disturbing the
  % existing application processes running on top of the database).
 
 \OMIT{
\item \textsf{Performance.} Finally, property graph databases are praised for
  their performance when working with highly-interconnected data. This is
  attributed to \textit{native graph data processing}, which may include, for
  instance, index-free adjacency.
  \angela{It is not clear how these key features of PG graph processing are related to schema (especially, performance). I guess we should not insist on performance/efficiency that much given that the paper has no experimental validation. }.
  }
% \item[Agility] PGs are close to conceptual data models which enables efficient translation from business concepts to database items and this provides high business responsiveness. 
% \item[Flexibility] PGs are typically schema-less (or even schema-last): new kinds of nodes and relationships can emerge as business needs are identified.
% \item[Efficiency] PGs enable native graph data processing algorithm (e.g., index-free adjacency).
\end{description}
Database schemas have a number of important functions that can be split into two general categories~\cite{BonifatiFGHOV19}. %\angela{These categories need to reflect what we already said in the Intro. }\george{I think they do now, with Filip's reintroduction in Section 1 of descriptive vs. prescriptive.  Question now is if we repeat this here or not.} \filip{It might be best to integrate as much of that discussion here somehow, and make the intro even lighter. It seems to fit better here.}
\begin{description}   
\item \textsf{Descriptive function.} Schemas provide a key to understanding the semantics of the data stored in a database. More precisely, a schema allows to construct a (mental) map between real-world information and structured data used to represent it. This knowledge is essential for any user and application that wishes to access and potentially modify the information stored in a database.
\item \textsf{Prescriptive function.} A schema is a contract between the database and its users that provides guarantees for reading from the database and limits the possible data manipulations that write to the database. To ensure that the contract is respected, a mandatory schema can be enforced by the database management system.
%\angela{How about: The database management system enforces a mandatory schema.}.    
\end{description}
Our primary objective is to develop a schema formalism for property graph databases that can effectively serve both descriptive and prescriptive roles, while also facilitating and possibly enhancing the strengths of property graph databases.
%Our main goal is to develop a schema formalism for property graph databases that fulfills both the descriptive and prescriptive roles while facilitating, and possibly enhancing, the strengths of property graph databases. 
Additionally, we aim at a formalism which enforces correct data modeling practices through its syntax. First, however, we discuss the meaning of two fundamental terms that are essential for defining schemas but are often confused.

\subsection{Types and Constraints}
The notions of \emph{type} and \emph{constraint} are two main building blocks in virtually any database schema formalism. When used sensibly, they enable the division of schematic information into self-contained fragments that correspond to real-world classes of objects (types) and pieces of knowledge about them (constraints). In general, there is no clear distinction between types and constraints, e.g., data value types are often considered as implicit (domain) constraints. 
%We employ the following uses of types and constraints throughout this paper:
We employ the following nomenclature throughout the paper.
%Below we outline the intended use for types and constraints that we employ throughout this paper.
\begin{description}
\item \textsf{Type} is a property that is assigned to elements (data values, nodes, edges) of a property graph database. Types group together similar elements that represent the same kind of real-world object and/or that share common properties, e.g., the set of applicable operations and the types of their results.
\item \textsf{Constraint} is a closed formula over a vocabulary that permits quantification over elements of the same type. The purpose of constraints is to impose limitations and to express semantic information about real-world objects.
\end{description}
Both types and constraints impose limitations (and provide guarantees) but types are of local nature while constraints are more global. More precisely, checking that an element has a given type should require only inspecting the element itself and possibly the immediately incident elements. On the other hand, validating constraints may require inspecting numerous database elements that need not even be directly connected. 
Consequently, types are typically verified statically whereas constraints are dynamically verified.
%\george{\textcolor{green}{@Sławek}, I reformulated this sentence based on Filip's comment (commented out in the next line).  Did I capture your intent here, as you wrote the original sentence?}\slawek{Yup, this looks good.}
%Consequently, types are often told \filip{What does 'told' mean in this context?} to allow static verification whereas constraints are told to require dynamic validation. 
%\angela{In Section 4, the part on constraints is very minimal, whereas here these two parts (types and constraints) seem to be at par. If I were a reader, I would be asking why. Shall we say somewhere that the constraints were studied in previous work?}\george{Agreed, Angela.  However, I don't think such a statement belongs in this section, as the focus here is purely on Requirements.  I think it would fit better in Section 1 and/or Section 4.}\slawek{I would simply indicate clearly in Section 4.5 that there has already been a whole paper on the topic}
\subsection{Requirements}

\subsubsection*{Property Graph Types}
%\angela{Why is this group of requirements called Agility?}\george{\textcolor{green}{@Angela}: see the def of Agility in section 2.1.  In particular, what Sławek and I are calling ``agility'' is the ability to nimbly work in terms of ``things and their relationships'' in the domain of discourse (i.e., the application domain), as this directly translates to ``nodes and their edges'' in the language of the property graph data model.  And, as a bare minimum, to support the PG data model, we need to support this conceptual modeling work in PG-Schema.  This translates to R1-R6.  Maybe we can better say ``Agility in data modeling: property graph types'' and ``Agility in data modeling: property graph constraints'', instead of the current headers for 2.3.1 and 2.3.2, resp.?}\slawek{We could also simply remove Agility from the subsection titles since we do indicate how this relates to property graph agility in the very first sentence.}  
The descriptive function of schemas can be particularly beneficial to the agility of property graph databases. Indeed, agility requires a good grasp of the correspondence between database objects and real-world entities, which is precisely the descriptive function of schemas. To communicate this correspondence efficiently, schemas should allow and even encourage a use of types that is consistent with how information about real-world objects is normally/typically divided into nodes, edges, and properties of a property graph database. Nodes are used to represent individual objects with all their attributes stored as node properties. 
\begin{itemize}
\item[\req{1}] {\sf Node types.} Schemas must allow defining types for nodes that specify their labels and properties.
\end{itemize}
Edges are used to represent relationships between objects of given kinds, and therefore should additionally specify node types of their endpoints. 
\begin{itemize}
\item[\req{2}] {\sf Edge types.} Schemas must allow defining types for edges that specify their labels and properties as well as the types  of incident nodes.
\end{itemize}
Naturally, schemas must also allow a great degree of expressiveness when describing the \emph{content} of nodes and edges, i.e., sets of properties and their values.
\begin{itemize}
\item[\req{3}] {\sf Content types.} Schemas must support a practical repertoire of data types in content types.
\end{itemize}

\subsubsection*{Property Graph Constraints}
%\angela{Why is this group of requirements called Agility?}\george{\textcolor{green}{@Angela}, please see my answer above to the first instance of this question}. 
To further ensure that schemas can properly fulfill the descriptive role and strengthen the agility of property graph databases, we additionally consider the data modeling power that a suitable schema formalism should have.  We deliberately target 
%\filip{Another 'deliberate' with different meaning very close. Better avoid it?} 
minimal data modeling capabilities and as a reference point we take the most basic variant of Entity-Relationship (ER) diagrams (see Section \ref{sec:discussion}), as the ultimate lower bound in the expressiveness of conceptual modelling languages.
%the standard class of ER diagrams~\cite{Chen1976}. \angela{The reference to ER models is not entirely clear to me at this point. Moreover, in Table 5, there are several variants. It would be good to justify this choice by anticipating what is in Table 2. }\filip{Might be enough to say that you take the most basic variant of ER diagram, as the ultimate lower bound in the expressiveness of conceptual modelling languages.} 
To that end, schemas, in addition to the previous requirements, must allow defining keys, which also provide the ability to define weak entities and functional (one-to-many) relationships.
\begin{itemize}
\item[\req{4}] {\sf Key constraints.} Schemas must allow specifying key constraints on sets of nodes or edges of a given type.
\end{itemize}
Schemas must allow for participation constraints, which mandate that nodes of a given type  participate in a relationship of a given type. 
\begin{itemize}
\item[\req{5}] {\sf Participation constraints.} Schemas must allow specifying participation constraints.
\end{itemize}
Finally, ER diagrams allow defining hierarchies of node types, a data modeling feature that is even more crucial for property graphs, where a single node may be an instance of multiple node types.
\begin{itemize}
\item[\req{6}] {\sf Type hierarchies.} Schemas must allow specifying type hierarchies.
\end{itemize}

\subsubsection*{Flexibility}
As we saw in Sections \ref{sec:introduction} and \ref{subsec:PG-and-schemas}, Property Graphs in practice are often popular in dynamic applications with volatile and evolving graph structures, 
where new kinds of objects are introduced following the evolving application demands. 
These typical scenarios require
support for flexible schema design 
across the full range between schema-first and schema-later, with evolvable and extensible schemas.

%and composable \stefania{this could be further detailed?}. 

\begin{itemize}
\item[\req{7}] {\sf Evolving data.} Schemas must allow defining node, edge, and content types with a finely-grained degree of flexibility in the face of evolving data.
\item[\req{8}] {\sf Compositionality.} Schemas must provide a fine-grained mechanism for compositions of compatible types of nodes and edges.
\end{itemize}

\subsubsection*{Usability}
Finally, schemas must be usable in practice. The basic requirements here are that the formalism must be implementable, and have well-defined semantics and a human-friendly declarative syntax. Furthermore, schemas must be easy to derive from graph instances and validation of graph instances with respect to schemas must be efficient.  These basic requirements are fundamental for the practical success of any schema solution, as we saw in Section \ref{sec:introduction}.

\begin{itemize}
\item[\req{9}] {\sf Schema generation.} There should be an intuitive easy-to-derive constraint-free schema for each property graph that can serve as a descriptive schema in case one is not specified.
\item[\req{10}] {\sf Syntax and semantics.} The schema language must have an intuitive declarative syntax and a well-defined semantics.
\item[\req{11}] {\sf Validation.} Schemas must allow efficient validation and validation error reporting.
\end{itemize}

%%% Local Variables:
%%% mode: latex
%%% TeX-master: "main"

%%% End:

\section{Data model}
\label{sec:preliminaries}

%data model based on LDBC PGS:JH-03r7 https://docs.google.com/document/d/13ECQZgqJfuBhHlCyxQfkNFjwG2jzwv5Md16ns9VyZWw

We assume countable sets $\LL$, $\KK$, and $\VV$ of \emph{labels}, \emph{property names (keys)}, and \emph{property values}. A \emph{record} with keys from $\KK$ and values from $\VV$ is a finite-domain partial function $o \colon \KK \pto \VV$ mapping keys to values.
We write $\RR$ for the set of all records.

\begin{definition}[Property Graph]
A \emph{property graph} is defined as a tuple $G = (N, E, \rho, \lambda, \pi)$ where:
\begin{itemize}[noitemsep,topsep=0pt,leftmargin=5mm]
\item 
$N$ is a finite set of nodes; 
\item 
$E$ is a finite set of edges 
such that $N \cap E = \emptyset$;
\item 
$\rho : E \rightarrow (N \times N)$ is a total function mapping edges 
to ordered pairs of nodes (the endpoints of the edge); 
\item 
$\lambda : (N \cup E) \rightarrow 2^{\LL}$ is a total function mapping nodes and edges 
to finite sets of labels (including the empty set);
\item 
$\pi : (N \cup E) \to \RR$ is a function mapping nodes and edges to  records. 
\end{itemize}
\end{definition}

For an edge $e \in E$ with $\rho_G(e) = (u,v)$, the nodes $u$ and $v$ are the \emph{endpoints} of $e$, where  $u$ is the \emph{source} and $v$ is the \emph{target} of $e$.
For an element $x \in N \cup E$, the record $\pi(x)$ collects all \emph{properties} of $x$ (key-value pairs) and is called the \emph{content} of $x$. 

%From a data modeling point of view, a node in a property graph represents an entity, an edge represents a relationship between entities.

%%a label represents a classification of an entity or relationship, and a property represents an attribute of an entity or relationship. 

%%% Local Variables:
%%% mode: latex
%%% TeX-master: "main"
%%% End:

\section{\pgschema}
\label{sec:proposal}

Most existing data definition languages for relational and semi\-structured data consist of two parts: \emph{types}, which define the basic topological structure of the data, and \emph{constraints}, which define data integrity.
Likewise, \pgschema consists of two parts.
The first part, \pgtypes, describes the shape of data and the types of its components such as nodes and edges, reflecting and extending work on SQL/PGQ schemas \cite{ISO9075,LDBC:OAEP:OAEP-2023-01}, 
Graph DDL in the openCypher Morpheus project for Apache Spark~\cite{morpheus}, and GQL graph types \cite{ISO39075,LDBC:OAEP:OAEP-2023-02}.\footnote{Note that in GQL, the term \emph{GQL-schema} refers not to a schema in our sense, but to a  dictionary of primary catalog objects such as graphs, graph types, or procedures.} It specifies 
\begin{itemize}
\item node types, describing the allowed combinations of labels and contents; 
\item edge types, describing the allowed combinations of labels, contents, and endpoint types; and
\item graph types, describing the types of nodes and edges present in the graph.
\end{itemize}
The second part describes constraints imposed on the typed data. Here, we propose a slight extension of the existing proposal called \pgkeys \cite{PGkeys}, which specifies integrity constraints such as keys and participation constraints, much like openCypher constraints \cite{LDBC:OAEP:OAEP-2023-03}. 
%https://s3.amazonaws.com/artifacts.opencypher.org/website/ocim1/slides/15-30+-+Language+Evolution-+Future+Features.pdf

This section starts with a guided tour of \pgtypes; the full syntax can be found in Figure~\ref{fig-grammar} and a parser is available on Zenodo \cite{grammar} (with a third-party web alternative \cite{dw2022}).
% available at \url{https://damianw27.github.io/pgs-grammar-check/}
We then define their semantics formally, provide a validation algorithm, and explain how \pgtypes interact with \pgkeys. 

\begin{figure*}[t]
\begin{center}
%\begin{minipage}{7cm}
\begin{lstlisting}
pgschema            ::= (createType ";"?)+ 
createType          ::= createNodeType | createEdgeType | createGraphType
createNodeType      ::= CREATE NODE TYPE (ABSTRACT)? nodeType
createEdgeType      ::= CREATE EDGE TYPE (ABSTRACT)? edgeType
createGraphType     ::= CREATE GRAPH TYPE graphType
graphType           ::= typeName graphTypeMode graphTypeImports? graphTypeElements
graphTypeMode       ::= STRICT | LOOSE
graphTypeImports    ::= IMPORTS typeName ("," typeName)*    
graphTypeElements   ::= "{" elementTypes? "}"
elementTypes        ::= elementType ("," elementType )*
elementType         ::= typeName | nodeType | edgeType
nodeType            ::= "(" typeName labelPropertySpec ")"
edgeType            ::= endpointType "-" middleType "->" endpointType
middleType          ::= "[" typeName labelPropertySpec "]"
endpointType        ::= "(" labelPropertySpec ")"
labelPropertySpec   ::= (":" labelSpec)? OPEN? propertySpec?
labelSpec           ::= "(" labelSpec ")"
                      | "[" labelSpec "]"
                      | labelSpec ( ( "|" | "&" ) labelSpec | "?" )
                      | label | typeName
propertySpec        ::= "{" (properties ("," OPEN)? | OPEN)? "}"
properties          ::= property ("," property)*
property            ::= (OPTIONAL)? key propertyType
typeName            ::= StringLiteral
label               ::= (*$l$*)  (*\:\textrm{ for } $l \in \LL$*)
key                 ::= (*$k$*)  (*\textrm{ for } $k \in \KK$*)
propertyType        ::= (*$b$*)  (*\textrm { for } $b \in \BB$*)
\end{lstlisting}
%\end{minipage}
\end{center}
\caption{Core productions of the \pgschema{} grammar with labels $\LL$, keys $\KK$, and base property types $\BB$}
\label{fig-grammar}
\end{figure*}

\subsection{\pgtypes by Example}
\label{ssec:syntax}

We first discuss the basic ingredients of \pgtypes (node types, edge types, and graph types) and then move on to more sophisticated aspects such as inheritance and abstract types. We use GQL's predefined data types like \inlinecode{DATE}, \inlinecode{STRING}, and \inlinecode{INT}. These are orthogonal to our proposal and could, in principle, be replaced by any other set of data types. Nevertheless, they take care of requirement~\req{3}. 

Generally, there are two main options for creating types in schemas. One can create \emph{open} types and \emph{closed} types. Both kinds of types are able to specify content that they \emph{require to be present}. The difference between the two is what they allow in addition to the explicitly mentioned content: closed types forbid any content that is not explicitly mentioned, whereas open types allow any such content.
%\bei{since graph types are included in the PG-TYPES above, does this open/closed concept apply to graph types too? Graph types are still with STRICT/LOOSE. Want to make sure it's intentional. }
%The idea behind closed types is that the type describes exactly the allowed content, i.e., the closed type describes which content is required and, additionally, everything that is not explicitly mentioned by the type definition is disallowed.\slawek{Can't optional properties be part of closed types? If yes, then this statement is potentially confusing. Perhaps stating that open types allow any properties in addition to those specified by the type whereas closed types prohibit any properties except for those explicitly specified }\wim{That's what the sentence said before Bei found it confusing -- and he was right. He wanted to know to which extent the types can also *require* stuff to be there. So I tried to make it a little bit more precise.}\slawek{I see, but open vs close is not about what it can require but what it can allow. Perhaps we  could say that the purpose of a node type is to specify properties and their types and to specify which of those are required and which are optional. Then move the discussion to open vs close?} 
%WIM: I gave it a shot. I did not mention optionality, because we haven't even introduced it yet.
Closed types are what we have in SQL, but also in programming languages such as C++ and Java. 
Open types are the default in JSON Schema.
%\footnote{A JSON Schema type is open by default and can be made closed by adding \texttt{"additionalProperties": false}.} 
%\stefania{maybe let's put a reference to Section 5 here, because we also mention this in 5.3?}
%WIM: @Stefania: I don't want to go too deep here, since the goal is just to explain quickly what open types are. I think that referring to one particular language in Sec 5 could be a bit distracting here. 
We provide both options here, and use the keyword \inlinecode{OPEN} to indicate the places where we use open types.
%To improve the usability of \pgtypes, we allow a form of \emph{multi-inheritance}. 
We use  declarative syntax closely aligned with the syntax of types in GQL. It adopts the evocative ASCII-art formatting \inlinecode{(\;)} for node types and \inlinecode{(\;)-[\;]->(\;)} for edge types, originating from Cypher~\cite{cypher}.

\subsubsection*{Base Node Types}

The most basic type is a node type. The following example specifies a node type for representing a person:
\begin{lstlisting}
  (personType: Person {name STRING, OPTIONAL birthday DATE})
\end{lstlisting}
It specifies a node of type \inlinecode{personType} with a label \inlinecode{Person}.
%In GQL, there are no special symbols that prefix type names. 
To distinguish type names from labels, we end type names with the suffix \texttt{Type}. 
%The ASCII art \inlinecode{(\,)} around the expression is meant to evoke the intuition that the label should be associated to a node. 
%In order to be able to distinguish easily type names from e.g., labels, we use the following naming convention: types names start with \inlinecode{n} followed by a capital letter, labels start with a capital letter, and property names are entirely lower case.\wim{It seems that someone is changing this. If so, please also update this description.}
%
By default, types are closed. That is, \inlinecode{Person} is the only allowed label. To permit nodes of type \inlinecode{personType} to have arbitrary additional labels, one should use the keyword \inlinecode{OPEN} and write 
\begin{lstlisting}
  (personType: Person OPEN {name STRING, OPTIONAL birthday DATE})
\end{lstlisting} 
In terms of properties, the type requires the node to have a property \inlinecode{name} of type \inlinecode{STRING}. Optionally, the node can have a property \inlinecode{birthday}. If it is present, it should have type \inlinecode{DATE}. No additional properties are allowed for this node type. Again, if we would like to allow them, we should write 
\begin{lstlisting}
  {name STRING, OPTIONAL birthday DATE, OPEN}
\end{lstlisting}
inside the definition.
More precisely, this content description  specifies that nodes should have \inlinecode{name} of type \inlinecode{STRING} and arbitrary additional properties. If the property \inlinecode{birthday} is present, its type should be \inlinecode{DATE}.
%It is important to 
Notice that the \inlinecode{OPEN} modifier applies independently to labels and properties: \inlinecode{OPEN} inside \inlinecode{\{...\}} applies to properties only and the occurrence outside applies to labels only.

Nodes in property graphs carry \emph{sets} of labels. In \pgtypes, we can associate multiple labels to a node type using the \inlinecode{\&}-operator:
\begin{lstlisting}
  (customerType: Person & Customer {name STRING, OPTIONAL since DATE})
\end{lstlisting}
The node type \inlinecode{customerType} requires nodes to carry both labels \inlinecode{Person} and \inlinecode{Customer}, and no other labels.
In general, we specify the allowed combinations of labels with a variant of \emph{label expressions} built from $\ell$ (labels) and $\ell?$ (optional labels) using operators $\&$ (and), $|$ (choice). Syntactically, these constitute a subset of label expressions used by GQL and SQL/PGQ for pattern matching in queries.
%\bei{do we want to define this more concretely / completely? (not that I'm advocating for it, just curious) For example, would label negation be useful?} 
% WIM: Bei, we don't know yet how to elegantly deal with negation. 
% We agree that it's interesting but it looks like it will spin off another paper.
We define their semantics in Section~\ref{ssec:formal}. 
%\bei{can the specific part of the section be called out? e.g. using subsection numbers? It's a little hard to locate at a glance.}
% WIM: Bei, it's a large chunk of the section (but maybe it wasn't written yet when you wrote this).
Intuitively, \inlinecode{A} \inlinecode{\&} \inlinecode{B?} would require \inlinecode{A} and additionally allow \inlinecode{B}; and \inlinecode{A} \inlinecode{|} \inlinecode{B} gives the choice between the label \inlinecode{A} or \inlinecode{B} (not allowing both). It is easy to define an \emph{inclusive or} \inlinecode{A}  \inlinecode{\textbackslash/} \inlinecode{B} as syntactic sugar for \inlinecode{A} \inlinecode{|} \inlinecode{B} \inlinecode{|} \inlinecode{(A} \inlinecode{\&} \inlinecode{B)}. 
A label expression can be accompanied with \inlinecode{OPEN} which, if specified, allows arbitrary additional labels. 

This part of \pgschema fulfils requirement \req{1}. Since we will introduce more advanced node types later using inheritance, we refer to the node types that we explained here as \emph{base node types}.

\subsubsection*{Base Edge Types} 

Let us define an edge type called \inlinecode{friendType}. 
%Similar to node types, we will start every edge type name with \inlinecode{e}, followed by a capital letter. 
Edges of type \inlinecode{friendType} carry the labels \inlinecode{Knows} and \inlinecode{Likes}, and connect two nodes of type \inlinecode{personType}. They are required to have a property  \inlinecode{since} of type \inlinecode{DATE}. The ASCII art \inlinecode{(\,)-[\,]->(\,)} indicates that we are talking about edges.
%\begin{lstlisting}
%(:personType)
%  -[friendType: Knows & Likes {since DATE}]->
%(:personType)
%\end{lstlisting}
\begin{lstlisting}
  (:personType)
    -[friendType: Knows & Likes {since DATE}]->
  (:personType)
\end{lstlisting}
%While the ASCII art alone distinguishes node types from edge types, it may not always be easy to see if a complex ASCII-art expression describes a node or edge. We propose to prefix type definitions with \inlinecode{NODE} or \inlinecode{EDGE} to enhance readability.
If one would like to be more liberal and allow \inlinecode{customerType} nodes on the ends of \inlinecode{friendType} edges, one could use the \inlinecode{|}-operator:
\begin{lstlisting}
  (:personType|customerType)
    -[friendType: Knows & Likes {since DATE}]->
  (:personType|customerType)
\end{lstlisting}
%\wim{The only reason why the above looks kind of nice is because nodes cannot be of type \inlinecode{personType} and \inlinecode{customerType} at the same time. If this would be possible, we'd have to write \inlinecode{personType|customerType|(personType\&customerType)}. I see that the "choose one" OR can be useful, but this example makes me wonder if the inclusive or isn't worth adding.}\filip{Sure it is. We just do not yet know how to do it in an elegant way... Future work :-)} 
One could be even more liberal and use \inlinecode{personType OPEN} to allow arbitrary labels and properties in addition to the material required by \inlinecode{personType}. 
%\wim{So a node can match \inlinecode{personType OPEN} without being of type \inlinecode{personType}. We could point this out but it's maybe a bit too subtle for here?} \filip{Is it? What do others think?} \wim{Notice that we haven't said anything at all about inheritance or combining existing types with other information at this point.} \filip{Oh, right. Maybe drop it then.}
This part of \pgschema fulfils requirement \req{2}.

\subsubsection*{Graph Types}
% WIM: It would be cool if we could put this early in the section because it makes it easy for the reader to get a quick grasp of how you can define a graph type using low-tech. Most users won't need inheritance.

A graph type combines node and edge types in one syntactic construct. It includes the types of the schema, as we will see here, but also the constraints, which we will see in Section~\ref{ssec:constraints}.
%\bei{can we unify the OPEN/CLOSED semantics and syntax for content, edge/node and graph types? When a graph type is marked as CLOSED, only the explicitly defined types are allowed, otherwise when marked as OPEN, additional types are also allowed?} \wim{Aligning syntax of course easy. But OPEN/CLOSED means something different from LOOSE/STRICT. Even with an OPEN declaration for labels, one can require certain labels to be present in the set. For OPEN/CLOSED records, it's similar. But with graph types, this is not true: LOOSE does not require nodes of a given type to be present in the graph. Not even STRICT can require that.} \bei{ Indeed. I agree with you Wim. I think it's probably safe to keep them as-is.} 
Here is an example:
%\begin{lstlisting}
%CREATE GRAPH TYPE fraudGraphType STRICT {
%  (personType: Person 
%               {name STRING, OPTIONAL birthday DATE}),
%  (customerType: Person & Customer 
%                 {name STRING, OPTIONAL since DATE}),
%  (suspiciousType: Suspicious OPEN
%                   {reason STRING, OPEN}),
%  (:personType|customerType)
%    -[friendType: Knows & Likes]->
%  (:personType|customerType)
%}
%\end{lstlisting}
\begin{lstlisting}
  CREATE GRAPH TYPE fraudGraphType STRICT {
    (personType: Person {name STRING, OPTIONAL birthday DATE}),
    (customerType: Person & Customer {name STRING, OPTIONAL since DATE}),
    (suspiciousType: Suspicious OPEN {reason STRING, OPEN}),
    (:personType|customerType)
      -[friendType: Knows & Likes]->
    (:personType|customerType)
  }
\end{lstlisting}
The graph type \inlinecode{fraudGraphType} contains three node types and one edge type. The keyword \inlinecode{STRICT} specifies how a property graph should be typed against the schema. It means that, for a graph $G$ to be valid \mbox{w.r.t.} \inlinecode{fraudGraphType}, it should be possible to assign at least one type within \inlinecode{fraudGraphType} to every node and every edge of $G$. The alternative, \inlinecode{LOOSE}, allows for \emph{partial validation}, addressing  \req{7}. Informally, it means that the validation process simply assigns types to as many nodes and edges in the graph as possible, but without the restriction that every node or edge should receive at least one type. We discuss this further in Section~\ref{ssec:formal}. 

We would like to point out the difference between open/closed element types and loose/strict graph types. Why do we use different terminology (and keywords) here? Element types work fundamentally differently from graph types. A node type of the form
\begin{lstlisting}
  (nodeType: Label {prop STRING, ...})
\end{lstlisting}
requires each node of type \inlinecode{nodeType} to have a property \inlinecode{prop}. A graph type such as \inlinecode{fraudGraphType} does not require nodes of type \inlinecode{customerType}. It merely requires that every node gets assigned \emph{some} node type declared in the graph type. Therefore, an open node type can require a given label to be present in a node, but a loose graph type cannot require a given element type to be present in the graph. 

%\textcolor{purple}{TODO}.
%\bei{What makes Graph Types different from node/edge types, that they have different semantics wrt OPEN/CLOSED? Why not Content, Node/Edge types behave the same as Graph types? I wonder if there is a deeper reason, which might be worth calling out in the paper? }

The example also shows the keyword \inlinecode{CREATE}. If a node or edge type is created as a catalog object, the declaration should likewise be preceded by \inlinecode{CREATE NODE TYPE} or \inlinecode{CREATE EDGE TYPE}, respectively. 
%\filip{Shall we make it \inlinecode{CREATE NODE TYPE} and \inlinecode{CREATE EDGE TYPE}, to make it consistent with \inlinecode{CREATE GRAPH TYPE}?} 
Node (and edge) types outside \inlinecode{CREATE GRAPH TYPE} statements should therefore always start with \inlinecode{CREATE}. If \inlinecode{personType} and \inlinecode{customerType} had been already created outside, one could define \inlinecode{fraudGraphType} more succinctly as follows.
%\begin{lstlisting}
%CREATE GRAPH TYPE fraudGraphType STRICT {
%  personType,   // import the type personType
%  customerType, // import the type customerType
%  (suspiciousType: Suspicious OPEN
%                   {reason STRING, OPEN}),
%  (:personType|customerType)
%    -[friendType: Knows & Likes]->
%  (:personType|customerType)
%}
%\end{lstlisting}
\begin{lstlisting}
  CREATE GRAPH TYPE fraudGraphType STRICT {
    personType,   // import the type personType
    customerType, // import the type customerType
    (suspiciousType: Suspicious OPEN {reason STRING, OPEN}),
    (:personType|customerType)
      -[friendType: Knows & Likes]->
    (:personType|customerType)
  }
\end{lstlisting}
This leads us to a subtle difference between simply \emph{referring} to a type that has been declared outside of the definition of a graph type, versus \emph{importing} such a type. By default, we are always allowed to refer to any type $t$ that is a catalog object. So, by omitting the import of \inlinecode{personType} and \inlinecode{customerType}, the edge type \inlinecode{friendType} would still be well-defined. However, by \emph{importing} $t$ we also allow objects in the graph type to be assigned the type $t$, which is important for the notion of \emph{validity} of a graph. When checking if a property graph $G$ is valid against \inlinecode{FraudGraphType}, one needs to be able to assign at least one type $t$ to each element of $G$ such that $t$ is either \emph{declared within} \inlinecode{FraudGraphType} or \emph{imported to} \inlinecode{FraudGraphType}.

\OMIT{
Within a graph type, one can \emph{refer} to types defined in the catalog; we did this for \inlinecode{\#Person} and \inlinecode{\#Customer}. Node (and edge) types can also be specified locally within a \emph{graph type}, in which case the keyword \inlinecode{CREATE} is not necessary. This is the case for \inlinecode{\#Suspicious}. If we wrote
\begin{lstlisting}
  CREATE NODE TYPE #Suspicious [...]
\end{lstlisting}
then we would be able to refer to \inlinecode{\#Suspicious} outside \inlinecode{\#FraudGraph}.\wim{Is this how we intend it?}\slawek{I think that we don't want to enter here in this discussion: I don't think that meta-data management (visibility, catalogue, etc.) is in the scope of the paper and perhaps it's better stay clear of it? }
}

\subsubsection*{Inheritance} 
Specifying contents for all relevant combinations of labels explicitly can be cumbersome and error-prone. We therefore 
allow reusing previously defined types in definitions of other types. Such reuse not only makes schemas more compact and modular, but also allows schema designers to follow a natural approach of classifying things as more general or more specialised, as is done in object-oriented modeling. With this mechanism, we fulfil requirement \req{6} (type hierarchies).

In the following example, the node type \inlinecode{employeeType} inherits labels and properties from \inlinecode{personType} and \inlinecode{salariedType}. 
\begin{lstlisting}
  (salariedType: Salaried {salary INT})
  (employeeType: personType & salariedType)
\end{lstlisting}
%The node type \inlinecode{employeeType} inherits labels and properties from \inlinecode{personType} and \inlinecode{salariedType}. 
That is, a node of type \inlinecode{employeeType} has the labels \inlinecode{Person} (inherited from \inlinecode{personType}) and \inlinecode{Salaried} (inherited from \inlinecode{salariedType}).
Its properties are \inlinecode{name} and optionally \inlinecode{birthday} (inherited from \inlinecode{personType}), as well as \inlinecode{salary} (inherited from  \inlinecode{salariedType}).
Note that inheritance automatically conflates properties that are compatible. If \inlinecode{salariedType} had a property \inlinecode{name}, then \inlinecode{employeeType} would only be well-defined if its \inlinecode{name} were a \inlinecode{STRING}. 

Similar to nodes, \pgtypes allow using edge types when specifying another edge type, which allows inheritance for edge types:
\begin{lstlisting}
  (:employeeType)
    -[buddyType: friendType {since DATE, casual BOOL}]->
  (:employeeType)
\end{lstlisting}
The edge type \inlinecode{buddyType} is an edge of type \inlinecode{friendType} but restricts the end nodes to be of type \inlinecode{employeeType}, i.e., end nodes also need to have a \inlinecode{salary} property and \inlinecode{Salaried} label. 
%\bei{can a few more words to be said, about the relationships between this new node restriction and the Friend edge type? For example do the new restriction have to be sub-types?}\domagoj{Added a sentence on this. Is it understandable now?} 
%\wim{Maybe we should say briefly how edge types are combined. My understanding is that each component is combined using $\oplus$. Don't know how to say that understandably yet. If someone wants to give it a shot (and check our compilation scheme to see if I'm right --- I didn't check the details), go for it!}
The type additionally requires the properties \inlinecode{since} of type \inlinecode{DATE} and \inlinecode{casual} of type \inlinecode{BOOL}. Notice that \inlinecode{since} is already required by \inlinecode{friendType}, so type \inlinecode{buddyType} would not change if we omitted it from the definition of \inlinecode{buddyType}. %\slawek{I would suggest extending this sentence with "[if we omitted it] in the type definition" or perhaps remove it from the definition and indicate that it is inherited} 
Intuitively, we can think  that inherited types collect all the property specifications of the parent types, and add the newly specified ones. If we declared \inlinecode{since} to be of a different type than \inlinecode{DATE}, the resulting edge type would be impossible to instantiate, and as such it would be redundant. The precise rules for how edge types and node types of endpoints are combined are in Section~\ref{ssec:formal}.

We also support \emph{graph inheritance}, which amounts to importing all node and edge types from one graph type to another graph type. For example,  by writing 
\begin{lstlisting}
  CREATE GRAPH TYPE fraudGraphType STRICT IMPORTS socialGraphType {...}
\end{lstlisting}
we import to \inlinecode{fraudGraphType} all types in \inlinecode{socialGraphType}.

\subsubsection*{Including Types in Label Expressions}
We can combine inheritance with adding new properties or labels. For instance, if we wrote 
%\filip{Something missing here...}\domagoj{I think the following was meant.}
\begin{lstlisting}
  (employeeType: personType & salariedType {birthday DATE})
\end{lstlisting}
then \inlinecode{birthday} would be a mandatory property in nodes of type \inlinecode{employeeType}, in addition to inherited properties \inlinecode{salary} and \inlinecode{name}.

Formally, we combine properties using the $\oplus$-operator, inspired by mixins \cite{BrachaC90} and explained in Section~\ref{ssec:formal}. Abstractly, if we define types
\begin{lstlisting}
  (xType: A & B {propertyA INT, propertyB INT})
  (yType: B & C {propertyB INT, propertyC INT})
  (zType: xType & yType)
\end{lstlisting}
then nodes of type \inlinecode{zType} have all labels \inlinecode{A}, \inlinecode{B}, and \inlinecode{C}, and all properties \inlinecode{propertyA}, \inlinecode{propertyB}, and \inlinecode{propertyC}. Since both \inlinecode{xType} and \inlinecode{yType} are closed, this means that a node can be of type \inlinecode{zType}, but not of type \inlinecode{xType} and not of type \inlinecode{yType}.

If both \inlinecode{xType} and \inlinecode{yType} were open types, i.e., declared their label sets and contents to be \inlinecode{OPEN}, then nodes of type \inlinecode{zType} would automatically fulfill \inlinecode{xType} and \inlinecode{yType}. That is, for open types, we support \emph{intersection types} via the operator \inlinecode{\&}. 
More generally, in the definitions of types we allow arbitrary expressions built from labels and previously defined types using operators \inlinecode{?}, \inlinecode{\&}, and \inlinecode{|}. 
% \bei{minor, feel free to just close: it then probably should not be call 'label' expression anymore but feel free to resolve this comment with no actions} 
Of course, declarations of node types should only refer to node types and similarly for edge types. Also, references should not be cyclic, as is standard in inheritance hierarchies.
%in object-oriented programming languages.  
Notice that using \inlinecode{|} we can define a \emph{union type}, which allows going beyond base types. For instance,
\begin{lstlisting}
  (aType: A {propertyA INT})
  (bType: B {propertyB INT})
  (cType: aType | bType)
\end{lstlisting}
creates a node type \inlinecode{cType} which either has the label \inlinecode{A} or \inlinecode{B}. 
However, \inlinecode{A} is only allowed to occur together with \inlinecode{propertyA} and \inlinecode{B} only with \inlinecode{propertyB} (but not \inlinecode{A} with \inlinecode{propertyB}). One cannot define \inlinecode{cType} as a base type, since base types always admit all combinations of matching label sets and matching property sets.
%\slawek{While reading the formal definition I've realized something about $|$, it's not a union but rather exclusive disjunction. In fact, suppose that A and B above have OPEN content types, then it is possible to have a node that satisfies both types A and B but it wont satisfy the type C. It's logical but also a bit disturbing and we might want to warn the reader about it.} \filip{With closed label sets, yes.  If both labels sets and contents were open, this would work as logical OR. We now say it explicitly the first time  we use |.}
%
Furthermore, if one of the base types were open, the derived type would automatically be open. This holds for both label openness and property openness. However, if both base types are closed, the derived type can be declared open (independently for labels and properties). 

The inclusion of types in label expressions makes the formalism highly compositional, fulfilling requirement \req{8}.

\OMIT{
Similar to inheritance in node types, we allow mixing labels and types. For example, the rule
\begin{lstlisting}
EDGE TYPE #FamilyFriend 
(#Person)-[#Friend & family]->(#Person)
  { since: DATE, CLOSED }
\end{lstlisting}
defines an edge of type \inlinecode{\#FamilyFriend}, which is the same as the edge type \inlinecode{\#Friend} but requires the edge to have an additional label \inlinecode{family}. The type \inlinecode{\#Familyfriend} could alternatively be written as follows.
\begin{lstlisting}
EDGE TYPE #FamilyFriend 
(#Person)-[family]->(#Person) & #Friend 
{ since: DATE, CLOSED }
\end{lstlisting} \filip{I have been thinking, maybe it should rather be \inlinecode{(\#Person)-[family]->(\#Person) \& ()-[\#Friend]->()} ? And then we could drop the explanation? Or maybe we do not need to talk about it at all?}
Since \inlinecode{\#Friend} is an edge type, the expression abbreviates
\begin{lstlisting}
(#Person)-[family]->(#Person) & 
(#Person)-[Knows & Likes]->(#Person)
\end{lstlisting}
} %End OMIT

\subsubsection*{Abstract Types} In some cases, one may want to declare a type as \emph{abstract}, which means that it cannot be directly instantiated. %\footnote{More precisely, while an element may fulfill the specification of an abstract type, we require for validity of the schema that it is also valid \mbox{w.r.t.} a non-abstract type.}\slawek{I don't think this foot note is necessary, the term "instantiable" has been introduced on intuitive level earlier and here it conveys sufficiently the meaning. If left, perhaps we should use the term "conforms to" rather than "valid w.r.t."}
\begin{lstlisting}
  ABSTRACT (salariedType {salary INT})
  (employeeType: personType & salariedType)
\end{lstlisting}
Notice that \inlinecode{salariedType} specifies nodes with no labels and a single property \inlinecode{salary}. On its own, this type may not be very useful, since we expect nodes with property \inlinecode{salary} to have labels and possibly other properties as well. Through inheritance from \inlinecode{salariedType}, the type \inlinecode{employeeType} matches nodes with all the labels and properties allowed in \inlinecode{personType}, plus an additional property \inlinecode{salary}.

\subsection{Formal Definition and Semantics}
\label{ssec:formal}

So far we have been talking about graph types by means of syntax. We now present a syntax-independent definition. Later we shall see how the two connect, thus providing the semantics for our declarative syntax, and fulfilling design requirement \req{10}.

\subsubsection*{Types and conformance} Recall from Section~\ref{sec:preliminaries} that $\LL$ is the set of labels, and $\RR$ the set of all possible records.
We define a \emph{formal base type} as a pair $(L,R)$, where $L \subseteq \LL$ and $R \subseteq \RR$. We write $\TT$ for the set of all formal base types. An element (a node or an edge) with label set $K$ and content $o$ \emph{conforms} to a formal base type $(L,R)$, if $K = L$ and $o \in R$. For the formal definition, we allow arbitrary subsets of $\RR$ to form base types. In the concrete syntax of the previous section, these will be given by record types; e.g., for the type \inlinecode{\{a INT, b STRING\}}, the set $R$ consists of all partial functions that map \inlinecode{a} to an integer and \inlinecode{b} to a string.

\begin{definition}
A \emph{formal graph type} is a tuple $S = (N_S,E_S,\nu_S,\eta_S)$ where
\begin{itemize}
\item $N_S$ and $E_S$ are disjoint finite sets of node and edge type names;
\item $\nu_S: N_S \to 2^\TT$ maps node type names to sets of formal base types; 
\item $\eta_S : E_S \to 2^{\TT \times \TT \times \TT}$ maps edge type names to sets of triples of formal base types: one for the source node, one for the edge itself, and one for the target node. 
\end{itemize}
\end{definition}
For brevity, we shall often refer to the elements of $N_S$ and $E_S$ as node and edge types, rather than node and edge type names. 
For dealing with \emph{strict} and \emph{loose} typing, we will use slightly different but connected notions, namely \emph{conformance} and \emph{typings}.
\begin{definition}
Let $G=(N_G,E_G, \lambda_G, \rho_G, \pi_G)$ be a property graph and $S = (N_S,E_S,\nu_S,\eta_S)$ be a formal graph type.
A node $v \in N_G$ \emph{conforms} to a node type $\tau \in N_S$ if it conforms to a formal base type in $\nu_S(\tau)$. 
An edge $e \in E_G$ \emph{conforms} to an edge type $\sigma \in E_S$ if for the pair $(v_1, v_2) = \rho_G(e)$ there is a triple $(t_1, t, t_2) \in \eta_S(\sigma)$ such that 
$v_1$ conforms to $t_1$, $e$ conforms to $t$, and
$v_2$ conforms  $t_2$.
A property graph $G$ \emph{conforms} to a formal graph type $S$ if every element in $G$ conforms to at least one type in $S$.
\end{definition}

The \emph{typing of $G$} \mbox{wrt.} $S$ is the mapping $\types : N_G\cup E_G \to 2^{N_S} \cup 2^{N_S}$ defined as follows for all $u\in N_G$ and $e \in E_G$:
\[ \types(u)  = \{\tau \in N_S \mid u \text{ conforms to } \tau\}\,, \qquad
    \types(e)  = \{\tau \in E_S \mid e \text{ conforms to } \tau\}\,.
\]
%\begin{align*}
%    \types(u) & = \{\tau \in N_S \mid u \text{ conforms to } \tau\}\,,\\
%    \types(e) & = \{\tau \in E_S \mid e \text{ conforms to } \tau\}\,.
%\end{align*}
%
%\begin{enumerate}
%\item $\types(u) = \{\tau \in N_S \mid u$ conforms to $\tau\}$ for each $u \in N_G$,
%\item $\types(e) = \{\tau \in E_S \mid e$ conforms to $\tau\}$ for each $e \in E_G$.
%\end{enumerate}
Hence, $G$ conforms to $S$ if $\types$ maps all nodes and edges to non-empty sets of types.

%\slawek{I feel that it might be useful to split here the subsection 4.2 into two separate: 1) formal graph types and their semantics, 2) schema compilation. I find myself often wondering about this section trying to find the definition of the semantics of formal schemas.}

\subsubsection*{Schema compilation}
We now explain how to interpret the syntax described in Section~\ref{ssec:syntax} in terms of formal graph types introduced above, thus providing the semantics for the syntax. This process, which we call \emph{schema compilation}, will effectively amount to unravelling and normalising all type definitions.

Let $T$ be a syntactically represented graph type. We shall define a corresponding formal graph type $S=(N_S, E_S, \nu_S, \eta_S)$. For $N_S$ and $E_S$ we take the sets of node and edge type names used in $T$. Because type definitions in $T$ are acyclic, we can use a bottom-up approach to unravel them. 

Consider a node type definition \inlinecode{($\tau$:$F$)} in $T$. Recall that $F$ is an expression built from labels $\ell$ and node type names $\sigma$ using operators $?$,  $\&$, and $|$, followed by an optional keyword \inlinecode{OPEN} and an optional  content description $r$. 
%(applied to subexpressions delimited with parentheses). 
Assume that 
$\nu_S$ is already defined over all node type names $\sigma$ used in $F$ (the base case is when $\tau$ is defined as a base node type).
The expression $F$ defines the family $\nsem{F} \subseteq \TT$ of formal base types allowed for type $\tau$. 
Intuitively speaking, $F$ describes how the allowed formal base types can be generated, starting from the simplest ones, much like a regular expression describes how to generate words.

Let  $t_\emptyset = (\emptyset,\{\bot\})$ and $t_\ell = (\{\ell\},\{\bot\})$, where $\bot$ stands for the empty record. These are the empty formal base type and the formal base type of a single label, with no content. We add content using content descriptions, which are record types written as
\[r=\{\, 
\text{[\inlinecode{OPTIONAL}] } 
k_1\: \basetype_1,\, \ldots,\, 
\text{[\inlinecode{OPTIONAL}] } 
k_n\:\basetype_n, \, \text{[\inlinecode{OPEN}]}\,\}\,\] 
where the square brackets  mean that the keywords \inlinecode{OPTIONAL} and \inlinecode{OPEN} are optional, the $k_i$s are keys from $\KK$, and the $\basetype_i$s are base property types such as \inlinecode{INT} or \inlinecode{DATE}. Let $\extbasetype_i$ be the extent of $\basetype_i$ (e.g., $\mathbb Z$ for \inlinecode{INT}). 
When the keyword \inlinecode{OPEN} is present, the semantics $\nsem{r}$ of $r$ is the set of all records $o \in \RR$ such that for all $i \leq n$, if $k_i \in \dom(o)$ then  $o(k_i) \in \extbasetype_i$, and $k_i$ must belong to $\dom(o)$ unless it is preceded by the keyword \inlinecode{OPTIONAL}. 
When the keyword \inlinecode{OPEN} is absent, we additionally require that $\dom(o) \subseteq \{k_1,\ldots,k_n\}$.

With these tools, we could define the semantics of node types with a single label and a content description. In order to handle more complex types we need a way to combine formal base types. We begin from records. 
We call records $o_1, o_2 \in \RR$ \emph{compatible} if $o_1(k)=o_2(k)$ for each $k\in \dom(o_1)\cap \dom(o_2)$. For compatible $o_1$ and $o_2$ we define their \emph{combination} $o_1 \oplus o_2$ as $(o_1 \oplus o_2) (k) = o_1(k)$ for $k\in\dom(o_1)$ and $(o_1 \oplus o_2) (k) = o_2(k)$ for $k\in\dom(o_2)\setminus\dom(o_1)$. For sets $O_1, O_2 \subseteq \RR$ we let $O_1 \oplus O_2$ be the set of all records of the form $o_1 \oplus o_2$ for compatible $o_1\in O_1$ and $o_2 \in O_2$. This operation is akin to the natural join  known from relational algebra.  The only difference is that in relational algebra columns are fixed for each relation, whereas a set of records may contain records with different sets of keys. We lift the $\oplus$ operator to formal base types by letting $(L_1,R_1)\oplus(L_2,R_2)=(L_1\cup L_2,R_1\oplus R_2)$. Note that in the absence of content, this amounts to taking the union of two sets of labels. 

Now we can define the semantics recursively for all subexpressions of $F$ as follows:
%%This disjunction operation eventually goes back but not for the submission version
%\wim{We're getting a lot of heat in Section 4.1 for defaulting to ``choose-one-or'' (Domagoj, Jan, Sławek). Would the addition of an inclusive or, defined as $\nsem{F_1} \lor \nsem{F_2} = \nsem{F_1} \cup \nsem{F_2} \cup (\nsem{F_1} \oplus \nsem{F_2} )$ be terrible? For closed types, it seems to do a reasonable thing. For open types, it's equivalent to $F_1 | F_2$? (Symbol $\lor$ just for illustration - concrete symbol to be determined. Could even be \inlinecode{\textbackslash /} for ``logical or''.)}
%
\OMIT{
\begin{align*}
\nsem{\ell}  & = \{ t_\ell \} \,,\\
\nsem{\sigma}  &= \nu_S(\sigma)\,,\\
\nsem{F_1?} &=  \nsem{F_1} \cup \big\{t_\emptyset\big\}\,,\\
\nsem{F_1 \mathbin{|} F_2}  & = \nsem{F_1} \cup \nsem{F_2} \,,\\
\nsem{F_1 \mathbin{\&} F_2} & = \big\{(L_1,R_1)\oplus (L_2,R_2) \bigm| (L_i,R_i)\in \nsem{F_i}  \text{ for } i=1,2\big\} \,,\\
\nsem{F_1 \text{ \inlinecode{OPEN}}} &= 
\big\{(L,R) \bigm| \exists\, L' \subseteq L \text{ such that } 
(L',R)\in \nsem{F_1} \big\}\,,\\
\nsem{F_1\, r} &= 
\big\{\big(L,R\oplus \nsem{r}\big) \bigm|  (L,R)\in \nsem{F_1}\big\}\,.
\end{align*}
} %% END OMIT
\[\nsem{\ell}   = \big\{ t_\ell \big\}\,, \qquad\qquad\qquad \nsem{\sigma}  = \nu_S(\sigma)\,,\]
\[\nsem{F_1?} =  \nsem{F_1} \cup \big\{t_\emptyset\big\}\,, \qquad
\nsem{F_1 \mathbin{|} F_2}   = \nsem{F_1} \cup \nsem{F_2} \,,\]
\[\nsem{F_1 \mathbin{\&} F_2}  = \big\{(L_1,R_1)\oplus (L_2,R_2) \bigm| (L_i,R_i)\in \nsem{F_i}  \text{ for } i=1,2\big\} \,,\]
\[\nsem{F_1 \text{ \inlinecode{OPEN}}} = 
\big\{(L,R) \bigm| \exists\, L' \subseteq L \text{ such that } 
(L',R)\in \nsem{F_1} \big\}\,,\]
\[\nsem{F_1\, r} = 
\big\{\big(L,R\oplus \nsem{r}\big) \bigm|  (L,R)\in \nsem{F_1}\big\}\,.
\]
With that, the semantics of $\tau$ is defined as
\[\nu_S(\tau) = \nsem{F}\,.\]

For edge types, we proceed in the same vein as for node types. Consider an edge type in $T$ defined as \inlinecode{(:$F_{\mathsf{src}}$)-[$\tau$:$F$]->(:$F_{\mathsf{tgt}}$)}. Expressions $F_{\mathsf{src}}$ and $F_{\mathsf{tgt}}$ specifying the source and target endpoints are interpreted as explained above. 
Expression $F$ defines the set $\esem{F} \subseteq \TT\times\TT\times\TT$ of triples of formal base types, according to the following rules: 
\OMIT{
\begin{align*}
\esem{\ell} & =  \big \{ \big(t_\emptyset,t_{\ell},t_\emptyset\big)\big\}\,,\\
\esem{ \sigma } &= \eta_S(\sigma)\,,\\
\esem{ F_1? } &=  \esem{ F_1 } \cup \big \{ \big(t_\emptyset,t_\emptyset,t_\emptyset\big)\big\}\,,\\
\esem{ F_1 \mathbin{|} F_2 } & = \esem{ F_1 } \cup \esem{ F_2 }\,,\\
\esem{ F_1 \mathbin{\&} F_2 } & = \esem{ F_1 } \oplus \esem{ F_2 }\,,\\
\esem{ F_1 \text{ \inlinecode{OPEN}}} &= 
\big\{\big(t_1,(L,R),t_2\big) \bigm| 
\exists\, L'\subseteq L \text{ s.t. }
\big(t_1,(L',R),t_2\big) \in \esem{F_1} \big\}\,,\\
\esem{F_1\, r} &= 
\big\{\big(t_1,(L,R\oplus \nsem{r}),t_2\big) \bigm|  \big(t_1,(L,R),t_2\big)\in \esem{F_1}\big\}\,,
\end{align*} 
}
\[ 
\esem{\ell}  =  \big \{ \big(t_\emptyset,t_{\ell},t_\emptyset\big)\big\}\,,\qquad
\esem{ \sigma } = \eta_S(\sigma)\,,\qquad
\esem{ F_1 \mathbin{|} F_2 }  = \esem{ F_1 } \cup \esem{ F_2 }\,,
\]
\[
\esem{ F_1? } =  \esem{ F_1 } \cup \big \{ \big(t_\emptyset,t_\emptyset,t_\emptyset\big)\big\}\,,
\qquad \quad \esem{ F_1 \mathbin{\&} F_2 }  = \esem{ F_1 } \oplus \esem{ F_2 }\,,
\]
\[
\esem{ F_1 \text{ \inlinecode{OPEN}}} = 
\big\{\big(t_1,(L,R),t_2\big) \bigm| 
\exists\, L'\subseteq L \text{ s.t. }
\big(t_1,(L',R),t_2\big) \in \esem{F_1} \big\}\,,
\]
\[
\esem{F_1\, r} = 
\big\{\big(t_1,(L,R\oplus \nsem{r}),t_2\big) \bigm|  \big(t_1,(L,R),t_2\big)\in \esem{F_1}\big\}\,,
\]
where the $\oplus$ operator is extended 
%to formal base types by $(L,R)\oplus (L',R')=(L\cup L',R\oplus R')$ 
%\slawek{this notation could be introduced and used in the semantics of label expressions}\filip{It had been. But Leonid says this way it is easier for the SIGMOD audience.} \leonid{Sort of. My initial approach to give an easy to understand semantics without R so then sem(F1 | F2) is just $\cup$ of Sem(F1) and Sem(F2). Now of course $\oplus$ gets there.}
%\leonid{Btw, didn't we decide to change $\oplus$ to $\otimes$?} \wim{We considered it but didn't decide it. For the SIGMOD crowd, using $\oplus$ for ``adding things'' may be even not so bad.} 
%and
to sets 
$Y_1,Y_2 \subseteq \TT\times\TT\times\TT$ of triples of formal base types by 
\[Y_1 \oplus Y_2=\big\{ \big(s_1 \oplus t_1, s \oplus t, s_2 \oplus t_2\big) \bigm| \big(s_1, s, s_2\big)\in Y_1, \big(t_1, t, t_2\big)\in Y_2\big\}\,.\]
With that, we define the semantics of the edge type $\tau$, specified by \inlinecode{(:$F_{\mathsf{src}}$)-[$\tau$:$F$]->(:$F_{\mathsf{tgt}}$)}, as 
\[ \eta_S(\tau) = \big(\nsem{ F_{\mathsf{src}} } \times\{t_\emptyset\}\times \nsem{ F_{\mathsf{tgt}} }\big)\oplus \esem{ F }\,. \]

After all type definitions in $T$ have been unravelled, as the last step, we restrict the sets $N_S$ and $E_S$ to type names whose definitions in $T$ are not preceded by the keyword \inlinecode{ABSTRACT}.  

%\wim{Maybe we should claim that $S_T$ can be obtained from $T$ in polynomial time? Say that it is ``easy'', straightforward induction, etc.?}

Having defined how a syntactically represented graph type $T$ corresponds to a formal graph type $S_T = (N_S,E_S,\nu_S,\eta_S)$, we can now explain validation of a property graph $G$ against \inlinecode{LOOSE} and \inlinecode{STRICT} graph types. In both cases, one begins by computing the typing of $G$ \mbox{w.r.t.} $S_T$. If $T$ is \inlinecode{LOOSE}, this is where the \emph{validation of $G$ against $T$} ends; the mapping $\types$ fully specifies which element of $G$ can be assigned which types from $T$. If $T$ is \inlinecode{STRICT}, we do one more step, namely we test if $\types$ assigns at least one type in $S_T$ to each element in $G$. If it does, we say that $G$ \emph{conforms to} $T$.

%For a \inlinecode{STRICT} graph type $T$, we say that $G$ \emph{is valid w.r.t.\ $S$} 
%For \inlinecode{LOOSE} graph types $T$, the validation process for a graph $G$ simply computes the mapping $\typing$. For \inlinecode{STRICT} graph types $T$, the validation process computes the mapping $\types$

%The semantics of a \inlinecode{STRICT} graph type $T$ can now be defined in terms of the corresponding formal graph type $S_T$ defined as above: a property graph $G$ has type $T$ if it conforms to the formal graph type $S_T$. For \inlinecode{LOOSE} graph types, rather than saying that a property graph $G$ conforms or not to a graph type $T$, we want to know which \emph{elements} of $G$ conform to $T$. Again, this can be defined in terms of the corresponding formal graph type $S_T$: a node in $G$ conforms to $T$ if it conforms to some node type in $S_T$, and similarly for edges. \filip{Introduce "typing". Clean up the use of "valid wrt type", "is of type". Also in the subsection below, add the partial thing.}

\subsection{Validation and Graph Type Generation}
\label{ssec:validation}

We now discuss how to validate a graph against a graph type. For a formal graph type, the process is straightforward: for each node type we identify the nodes that conform to it, and we use this information to identify for every edge type the set of edges that conform to it. The validation for general graph types, defined with the syntax in Section~\ref{ssec:syntax}, can be accomplished efficiently with an analogous procedure thanks to the mathematical simplicity of the schema compilation rules in Section~\ref{ssec:formal}. More importantly, 
%it is very much conceivable that 
such a validation procedure can be implemented in a reasonably expressive graph query language. In essence, such a language would need to support standard set operations and would need to allow  identifying nodes and edges based on their labels, property names, and property value types. Consequently, the proposed graph schema formalism satisfies requirement \req{11}.

We also propose a simple method of generating a graph type for a given property graph. For every node, we introduce an anonymous node type that fits precisely its set of labels and properties, we gather those types into a set, and finally generate their names. Using anonymous types allows to eliminate repetitions of syntactically identical types easily. We next apply an analogous procedure for edges remembering to use previously identified node types of the endpoints. Finally, we group types of all elements into the resulting graph type. This shows the satisfaction of requirement \req{9}. One might wish for  a more refined method of schema generation, but those fall into the domain of schema inference~\cite{BonifatiDMGJLP22, GrozLSW22}, which is out of the scope of this paper and left as future work.

%\wim{There's some stuff here in latex comment that we may resurrect. It can help saying formally what we mean with \emph{typing}.}
%We show that typing against a graph type $S = (N_S, E_S, \nu_S, \eta_S)$ can always be done efficiently. The algorithm first computes,
%\begin{enumerate}
%\item for each node $u$ of $G$, the set of types $\tau_u := \{\tau \in N_S \mid u$ conforms to $\tau\}$ and,
%\item for each edge $e$ of $G$, compute the set of types $\tau_e := \{\tau \in E_S \mid e$ conforms to $\tau\}$.
%\end{enumerate}
%We refer to the above sets of types as the \emph{typing of $G$ w.r.t.\ $S$}. 
%If $S$ is \inlinecode{LOOSE}, the typing of $G$ w.r.t.\ $S$ already provides everything we need. If $S$ is \inlinecode{STRICT}, we additionally need to heck if $\tau_\ell$ is non-empty for every element $\ell$ of $G$.

\subsection{Adding Constraints}
\label{ssec:constraints}

Our focus until now was on \emph{types} in \pgschema. The other crucial aspect of \pgschema is its \emph{constraints}. To this end, we leverage existing work on \emph{keys for property graphs} \cite{PGkeys}, called \pgkeys. Despite their name, \pgkeys go beyond the capability of expressing key constraints. Statements in \pgkeys are of the form
\begin{lstlisting}[mathescape=true]
  FOR $p(x)$ <qualifier> $q(x,\bar y)$(*\comma*)
\end{lstlisting}
where \inlinecode{<qualifier>} specifies  the kind of constraint that is being expressed and consists of combinations of \inlinecode{EXCLUSIVE}, \inlinecode{MANDATORY}, and \inlinecode{SINGLETON}. Both $p(x)$ and $q(x,\bar y)$ are queries. 
For instance, if we want to express that, for every output $x$ of $p(x)$ there should be at least one tuple $\bar y = (y_1, y_2, \dots, y_n)$ that satisfies $q(x, \bar y)$, we write \inlinecode{FOR} $p(x)$ \inlinecode{MANDATORY} $q(x,y_1,\ldots,y_n)$. \inlinecode{SINGLETON} would mean that there should be at most one such $\bar y$ for each $x$, and \inlinecode{EXCLUSIVE} that no $\bar y$ should be shared by two different values of $\bar x$. Inside the queries, we can use the keyword \inlinecode{WITHIN} to make clear what the output of the queries is, i.e., what we want to be \inlinecode{EXCLUSIVE}, etc.
%\slawek{The meaning of the keyword WITHIN is not explained; it is however used in the examples.}
%keywords are explained in \cite{PGkeys}.
%\wim{WITHIN is now mentioned.}

%The keywords \texttt{EXCLUSIVE}, \texttt{MANDATORY}, and
%\texttt{SINGLETON} indicate which assertions the \pgkey makes:
%\begin{description}
%\item[{\tt EXCLUSIVE}] --- no two targets can have the same key value;
%\item[{\tt MANDATORY}] --- for each target there is at least one key value;
%\item[{\tt SINGLETON}] --- for each target there is at most one key value.
%\end{description}

In \pgschema, we slightly extend the syntax of \pgkeys by allowing the constraints to refer to a \emph{type name} at each point where \pgkeys allows a label. The semantics of the resulting expression is that \emph{a type name $t$ matches every node that conforms to $t$}.
%\wim{My feel is that the previous sentence already fully specifies how \pgkeys work with type names. Type tests work exactly in the same way as label tests in the query language and a type test matches if the node conforms to the type. This is the same as saying that constraints are evaluated post-validation, i.e., after types have been assigned to all nodes and edges in the graph.}

\OMIT{ %% PREVIOUS VERSION
\begin{lstlisting}
CREATE GRAPH TYPE socialGraphType STRICT {
  ...
  (:Person)
    -[bestFriendType: friendType & bestie]->(:Person)
  FOR (x:Person)
    SINGLETON y WITHIN (x)-[y::bestFriendType]->()
}
\end{lstlisting}

In this example, the edge type \inlinecode{bestFriendType} is defined as the edge type \inlinecode{friendType}, but additionally requires the label \inlinecode{bestie} on the edge. In our toy example, each person is allowed to have at most one best friend, which is captured by the \pgkey at the end of the schema.  Notice the difference between \inlinecode{x:Person} and \inlinecode{y::bestFriendType}. The former is a GQL label expression, stating that $x$ should have a label \inlinecode{Person}, and the latter states that $y$ should conform to the type \inlinecode{bestFriendType}.
}

Consider the following code snippet, describing a graph with two kinds of nodes, persons (\inlinecode{personType}) and customers (\inlinecode{customerType}), and friend-edges between persons (the edge type \inlinecode{friendType}, requiring labels \inlinecode{Knows} and \inlinecode{Likes} and allowing label \inlinecode{Bestie} on the edge). 
\begin{lstlisting}
  CREATE GRAPH TYPE socialGraphType STRICT {
    (personType: Person {name STRING, id INT}),
    (customerType: Customer {id INT}),
    (:personType)
      -[friendType: Knows & Likes & Bestie?]->
    (:personType),
    // Constraints
    FOR (x:personType) 
      EXCLUSIVE MANDATORY SINGLETON x.id,
    FOR (x:customerType) 
      MANDATORY y.id WITHIN (y:personType) WHERE y.id = x.id,
    FOR (x:personType) 
      SINGLETON y WITHIN (x)-[y: friendType & Bestie]->()
  }
\end{lstlisting}
Apart from type declarations, the graph type also has three \pgkey constraints. The first expresses that the value of the property \inlinecode{id} should be a key for nodes of type \inlinecode{personType}. The second \pgkey expresses that every \inlinecode{id} value of a customer should be an \inlinecode{id} of a person, which is a foreign key. \pgkeys (and therefore \pgschema) can therefore handle \emph{key and foreign key constraints} (requirement \textbf{R4}).
The third \pgkey expresses that each person is allowed to have at most one best friend. Notice that \inlinecode{y:~friendType \& Bestie} means that $y$ should have label \inlinecode{Bestie} and it should conform to the type \inlinecode{friendType}. If we wrote
\inlinecode{MANDATORY y WITHIN (x)-[y:friendType]->()} in the second constraint, it would express that each person participates in the \inlinecode{friendType} relation, i.e., it expresses a participation constraint (requirement \textbf{R5}). 
% Hence, \pgkeys (and therefore \pgschema) can handle \emph{participation constraints} (R5). \filip{This is more like a foreign key. I think what Sławek means by participation constraints is just that nodes of certain types must have some outgoing edges - participate in some relationships.}
\pgkeys are also powerful enough to express  SQL-style \inlinecode{CHECK} constraints, such as
\begin{lstlisting}[escapechar=@]
  FOR (x:salariedType) 
    MANDATORY x.salary >= 0@\,,@   
\end{lstlisting}
 or denial constraints, such as
 \begin{lstlisting}[escapechar=@]
  FOR (x:customerType) 
    MANDATORY (x:!employeeType)@\,,@   
\end{lstlisting}
where \inlinecode{!} denotes negation. 

The semantics of \pgkeys is the same in loose and strict graph types. In particular, constraints in loose graph types are not trivially satisfied and can be useful. For instance, by changing \inlinecode{STRICT} to \inlinecode{LOOSE} in the  graph type above, we allow nodes that are neither of type \inlinecode{personType}, nor of type \inlinecode{customerType}, but \inlinecode{id} must still be a key for all \inlinecode{personType} nodes.

Concerning validation, notice that the typing of a property graph $G$ w.r.t.\ a graph type $S$ can be computed efficiently. Once we have this typing, \pgkeys can be evaluated as in the original paper~\cite{PGkeys}, using types as if they were labels in the graph. 

We conclude this section with the observation that \pgschema  satisfies all requirements identified in Section~\ref{sec:requirements}.

%Add a paragraph about validation: Refer to the previous paper recalling that these constraints can be rewritten into queries that select target values for which constraints are violated; this allows to report local validation errors as postulated in the requirements. 

%%% THIS FILE USED TO BE CALLED 5-discussion_updated.tex. 
%%% I renamed it and moved the old 5-discussion.tex to the attic.
\section{Relationship to Other Paradigms}
\label{sec:discussion}

We now compare \pgschema with existing schema formalisms
%, listed in Section~\ref{ssec:existingschemas}, 
and with existing graph schema technologies.  We consider a wide range of formalisms. First, we consider \emph{conceptual data models} such as the Entity-Relationship Model and its variants.
Second, we discuss graph schemas for RDF graphs stemming from the \emph{Semantic Web} setting.
Third, we overview schema languages for \emph{tree-structured data} formalisms such as XML and JSON.
A detailed description of these existing schema formalisms is omitted for space reasons and can be 
found in an external technical report~\cite{pgsarxiv}. 

In order to perform this comparison, we define several features in Section~\ref{ssec:feat}, and discuss their support in Section~\ref{ssec:support}.
Finally, we propose potential \pgschema extensions in Section~\ref{ssec:extensions}.

%In this section we compare \pgschema to other schema formalisms. We first present these formalisms, then discuss the features we will consider and finally how they compare on these features. Finally we discuss possible extensions of \pgschema.

%\domagoj{Meta comment for the section after reading through it a few times: First, I think 5.1 is great, but there will be some repetitiveness with 5.3. Would it make sense to start with what is now 5.2, and then merge 5.1 and 5.3? If space is not an issue, I would still prefer the current configuration though.}
%\stefania{I will try to wrap up the Tree-structured data and Existing graph technologies parts in 5.3 asap, so that you have a better idea. I suspect that there will not be a lot of overlap bc we strictly refer to Table 2.}

%\subsection{Existing Graph Schema Formalisms}\label{ssec:existingschemas}

%\domagoj{I would suggest integrating this into the intro to the section, and this subsection can disappear?}

\subsection{Existing Graph Schema Features}\label{ssec:feat}

We briefly describe the main features used to compare state-of-the-art graph schema languages in Table~\ref{tab:featureTable}. %We group these into: type features (Section~\ref{sssec:tf}), constraint features (Section~\ref{sssec:cf}), and schema features (Section~\ref{sssec:sf}).
We group these into: type features, constraint features, and schema features. %, as explained next.

%We group the features in (1) type features, (2) constraint features and (3) schema features.
%\textcolor{purple}{The following can and should be formatted more compactly, but is left for the moment like this so it is easier to lookup for ourselves.}

\subsubsection*{Type features}\label{sssec:tf} 

We considered: \textbf{(PDT)} the number of built-in primitive data types, \textbf{(UIT)} type constructors for union and intersection types, \textbf{(TH)} type hierarchies, \textbf{(AT)} abstract types, \textbf{(OCT)} open and closed types, \textbf{(EP)} edge properties, \textbf{(MOP)} mandatory and optional properties, \textbf{(CPT)} complex nested property types consisting of nested collection types,  and \textbf{(RC)} range constraints.

\subsubsection*{Constraint features}\label{sssec:cf} 
We examined: \textbf{(KC)} key constraints, \textbf{(MP)} man\-datory participation of certain types of nodes in certain types of edges, \textbf{(CC)} cardinality constraints for such participation, and \textbf{(BRC)} properties of binary relations defined by certain edges, such as (ir)reflexivity, (in)transitivity, (a)cyclicity, (a/anti)symmetry, etc.

\subsubsection*{Schema features}
\label{sssec:sf} 

We assessed: \textbf{(TV)} if validation is tractable,
% \textbf{(PC)} if partial conformance can be allowed, 
\textbf{(ISP)} if introspection is possible, i.e., the schema can be queried like a graph instance, and finallly \textbf{(SFPX)} if the schema can be specified to be (1) \emph{first}, and subsequently enforcing it for all instances, (2) \emph{partial}, allowing some of its components to be descriptive (e.g., the element types) and some to be prescriptive (e.g., the constraints), or (3) \emph{flexible}, creating and updating instances in an unconstrained manner while possibly maintaining a descriptive schema.

\subsection{Support of the Features}
\label{ssec:support}
% \textcolor{purple}{Present here the matrix~\ref{tab:featureTable} and describe for each formalism if and how they support the features.}

We comment on some of the differences between \pgschema and the existing state-of-the-art graph schema formalisms and systems from Table~\ref{tab:featureTable}, with a focus on existing graph technologies. %\domagoj{Please explain what numbers in PDT column mean.}

%% was [ht]  
% \begin{table*}[p]
\begin{table*}[ht]
% \begin{adjustbox}{width=\textwidth} % this shrinks and rotates the table to fit the page
% \begin{minipage}{\textwidth}
%  \medskip
%  \begin{centering}
%
% the following settings make \hdashline produce dotted lines
\setlength\dashlinedash{0.2pt}
\setlength\dashlinegap{1.5pt}
\setlength\arrayrulewidth{0.3pt}
% \newcolumntype{g}{>{\cellcolor[rgb]{0.8, 0.1, 0.4}}c}
\caption{Overview of the features supported by state-of-the-art graph schema formalisms}
\label{tab:featureTable}
\scriptsize
   \textbf{PDT} = \emph{Primitive Data Types},
   \textbf{UIT} = \emph{Union and Intersection Types},
   \textbf{TH} = \emph{Type Hierarchy},
   \textbf{AT} = \emph{Abstract Types},
   \textbf{OCT} = \emph{Open/Closed Types},
   
   \textbf{EP} = \emph{Edge Properties},
   \textbf{MOP} = \emph{Mandatory/Optional Properties},
   \textbf{CPT} = \emph{Complex Property Types},
   \textbf{RC} = \emph{Range Constraints},
   
   \textbf{KC} = \emph{Key Constraints},
   \textbf{MP} = \emph{Mandatory Participation},   
   \textbf{CC} = \emph{Cardinality Constraints},
   \textbf{BRC} = \emph{Binary-Relation Constraints},
   
   \textbf{TV} = \emph{Tractable Validation},
%   \textbf{PC} = \emph{Partial Conformance},
   \textbf{IS} = \emph{Introspection},
   \textbf{SFPX} = \emph{Schema First/Partial/fleXible}
\\
\medskip
\begin{adjustbox}{max width=\textwidth} % this shrinks the table to fit the page, but only if necesary
% \begin{adjustbox}{max width=0.725\textheight, rotate=-90} % this shrinks the table to fit the page, but only if necesary
\setlength{\tabcolsep}{2pt}
  \begin{tabular}{@{}>{\bfseries}l@{\hskip 2pt} c c c c c c c c c @{\hskip 17pt} c c c c@{\hskip 15pt} c c c@{}} \toprule
  & \bf PDT & \bf UIT  & \bf TH    & \bf AT    & \bf OCT & \bf EP & \bf MOP & \bf CPT & \bf RC & \bf KC & \bf MP & \bf CC & \bf BRC & \bf TV & \bf IS & \bf SFPX \\ \midrule
    Chen ER \cite{Chen1976}  & \qual{\lacks}  & \lacks & \lacks & \lacks & c   & n/e & m   & \lacks & \qual{\lacks} & \qual{\has} & \has & \lacks  & \lacks & \qual{\lacks}&\lacks & f \\ \hdashline
    Extended ER \cite{Thalheim_extended_2018}  & \qual{\lacks}  & n/e   & n/e/p & n/e   & c   & n/e & m/o & \has & \qual{\lacks} & \has & \has & \has & \lacks & \qual{\lacks}&\lacks & f \\ \hdashline
    Enhanced ER \cite{Elmasri_fundamentals_2015}  & \qual{\lacks}  & n/e   & n/e  & n     & c   & n/e & m/o & \has & \qual{\lacks} & \qual{\has} & \has & \has & \lacks & \qual{\lacks}&\lacks & f \\ \hdashline
    ORM2 \cite{halpin2015}      & \qual{\lacks} & n/e/p  & n/e/p & n/e   & c   & n/e & m/o & \qual{\lacks} & \has & \has & \has & \has & \has & \qual{\lacks}&\lacks & f \\ \hdashline
    UML Class Diagrams \cite{Fowler2003} & \qual{5}  & n & n  & n     & c   & n/e & m/o & \qual{\lacks} & \has & \has & \has & \has & \qual{\has} & \qual{\lacks}&\lacks & f \\ \midrule
    RDFS \cite{Brickley2014RS}  & 34        & \lacks  &  n/e/p & \lacks  & o & \qual{\lacks} & \qual{o} & \lacks  & \lacks & \lacks & \lacks & \lacks & \lacks & \has&\has & f/\qual{p}/x \\ \hdashline
    OWL \cite{Hitzler2012OWO} &  \qual{33} & n/e/p     &  n/e/p & \lacks  & o  & n & \qual{m}/\qual{o} & \lacks & \has   & \has & \qual{\lacks} & \has & \has & \qual{\has}&\has & f/\qual{p}/x \\ \hdashline  
    SHACL \cite{Knublauch2017SCL,DBLP:conf/semweb/CormanRS18} & 34 & n/e/p & n/e/p & \lacks &  o/c   & n & m/o & \lacks   & \has   & \qual{\has}  & \has & \has & \lacks & \qual{\has} &\has & f/p/x \\ \hdashline
    ShEx \cite{Baker2019,Staworko2015complexity} & 34 & n/e/p & n/e/p & \lacks  &  o/c  & n & m/o & \lacks   & \has   & \qual{\has} & \has & \has & \lacks & \qual{\has} &\has & f/p/x \\ \midrule
    %ReSh     & \tbd  & \tbd     & \tbd     & \tbd     & \tbd   & \tbd   & \tbd   & \tbd & \tbd         & \tbd & \tbd & \tbd & \tbd & \tbd & \tbd & \tbd & \tbd \\ \hdashline \hdashline
    DTD \cite{Yergeau2008EML}      & 6  & \qual{n}  & \lacks  &  \lacks     &  o/c  & n   & m/o   & \qual{\lacks} & \lacks    & \qual{\lacks} & \has & \lacks & \lacks & \has& \qual{\has} & f/x \\ \hdashline
    JSON Schema \cite{Wright2020JS} & 6  & n/e  & n/e/p      &   n  & o/c   & n   & m/o & \has  & \has  & \qual{\has} & \has & \has & \lacks & \has&\has & f/x \\ \hdashline
    RELAX NG \cite{ISO19757-2} & \qual{2}  & n  & n/e/p      & n    & o/c   & n   & m/o   & \has & \has & \qual{\lacks} & \has  & \lacks & \lacks & \has&\has  & f/x \\ \hdashline
    XML Schema \cite{SperbergMcQueen2012WXS} & \qual{47}  & n    & n/e/p      & n    & o/c   & n   & m/o   & \has & \has         & \has & \has& \has & \lacks & \has&\has & f/x \\ \midrule
    GraphQL SDL \cite{GraphQL2018,Hartig2018semantics, GraphQLforPG} & 5  & n/\qual{e}  & n/e     & n/e     &  c   & n/e   & m/o & \has  & \qual{\has} & \has & \has & \lacks & \lacks & \qual{\lacks}&\has & f/\qual{p} \\ \hdashline
    openCypher Schema \cite{BonifatiFGHOV19} & [oC]  & [n]  & n/e     & n/e     &  c   & n/e   & m & \has  & \lacks & \has & \has & \lacks & \lacks & \has & \has & p/x \\ \hdashline
    SQL/PGQ \cite{ISO9075} & [SQL]  & \lacks     & \qual{n}/\qual{e}  & n/e     & c   & n/e & m/o  & \has   & \qual{\has} & \has & \has & \qual{\lacks} & \lacks & \has & \lacks & f \\ \hdashline
    GQL \cite{ISO39075}  & [-]  & -   & - & - & \qual{o}/c & n/e & m/o & \lacks & \lacks & \lacks & \lacks & \lacks & \lacks & \has & \lacks & f/x \\ 
%    ProGS \cite{Seifer21} (\textcolor{red}{TODO}) &  &  &  &  &     &  &  &   &   & & & & & &  \\    
%    SHACL values: & 34 & n/e/p & n/e/p & \lacks &  o/c   & n & m/o & \lacks   & \has   & \qual{\has}  & \has & \has & \lacks & \qual{\has} &\has & f/p/x \\    
    \midrule
    AgensGraph \cite{AgensGDB} & \qual{\lacks}  & \lacks & n/e/p & \lacks & o & \qual{n}/\qual{e} & m/o & \qual{\has} & \qual{\has} & \qual{\has} & \has & \lacks & \lacks & \qual{\has} & \qual{\has} & f/\qual{p}/x \\ \hdashline
    ArangoDB \cite{arango:website} & 6  & \qual{n}/\qual{e} & n/e/p     & n     & o/c  & n/e & m/o   & \has   & \qual{\has}         & \qual{\has} & \has & \qual{\has} & \lacks & \qual{\has} & \has & f/x \\ \hdashline
    DataStax \cite{datastax:website} & \qual{25}  & \lacks     & \lacks    & \lacks    & \qual{o}   & n/e & m/o   & \has   & \qual{\has}         & \has & \has & \lacks & \lacks & \unkn & \has & f/\qual{p}/x \\ \hdashline
    JanusGraph \cite{JanusDB}     & 12  & \lacks   & \lacks & \lacks   & \qual{o}   & n/e & m/o  & \has  & \lacks      & \has & \has & \has & \lacks & \has & \qual{\has} & f/\qual{p}/x \\ \hdashline
    Nebula Graph/nGQL \cite{NebulaDB}     & 5  & \lacks     & \lacks     & \lacks     & c  & n/e & m/o   & \lacks   & \lacks     & \lacks & \lacks & \lacks & \lacks & \qual{\lacks} & \has  & f \\ \hdashline
    Neo4j \cite{Neo4jDB}    & 11  & \qual{n} & \qual{\lacks}     & \qual{\lacks}    & o   & n/e   & m/o   & \qual{\has} & \lacks         & \has & \lacks & \qual{\lacks} & \lacks & \lacks & \has & p/x \\ \hdashline 
    Oracle/PGQL \cite{OracleSGDB}    & 11  & \qual{n}/\qual{e} & \lacks & \lacks & c & n/e & \qual{m}/\qual{o} & \lacks & \qual{\has} & \qual{\has} & \qual{\has} & \qual{\lacks} & \lacks & \qual{\has} & \has & f/\qual{p}/x \\ \hdashline
    OrientDB/SQL \cite{OrientDB} & 23  & \qual{n}/\qual{e} & \qual{n}/\qual{e} & \qual{n}/\qual{e} & o/c & n/e & \qual{m}/\qual{o} & \has & \qual{\has} & \qual{\has} & \qual{\has} & \qual{\lacks} & \lacks & \qual{\has} & \qual{\has} & f/\qual{p} \\ \hdashline
    Sparksee \cite{SparkseeDB} & 8  & \lacks & \lacks & \lacks & c   & n/e & m/o   & \qual{\has}   & \lacks         & \has & \has & \lacks & \lacks & \qual{\lacks} & \qual{\has} & f \\ \hdashline
    TigerGraph/GSQL \cite{TigerGraphDB}    & 8  & \qual{n}/\qual{e} & \lacks & \qual{\lacks}    & c  & n/e   & \qual{m}/\qual{o}  & \has  & \qual{\has} & \has & \lacks & \qual{\lacks} & \qual{\has} & \qual{\has} & \has & f \\ \hdashline
    TypeDB/TypeQL \cite{TypeDB} & 5  & \lacks & n/e/p & n/e/p & c & n/e & m & \has & \qual{\has} & \has & \has & \qual{\lacks} & \has & \qual{\lacks} & \has & f \\ \midrule
    \pgschema & [-]  & n/e   & n/e/\qual{\lacks} & n/e/\qual{\lacks}   & o/c & n/e & m/o & \qual{\lacks} & \qual{\lacks} & \has & \has & \qual{\lacks} & \qual{\lacks} & \has & \lacks & f/p/x \\ \bottomrule 
\end{tabular}
\end{adjustbox}
%  \end{centering}
 % \small 
 % \textcolor{purple}{
 %   \textbf{Working Legenda (to be removed in final version):} `\tbd' = \emph{yet to be determined}, \unsure{$x$} = \emph{mark $x$ is provisional and needs to be verified}
 % }
 \smallskip \\ 
   \textbf{Legend:} `\has' = \emph{supported}, `\lacks' = \emph{not supported}, `\unkn' = \emph{unknown}, \qual{$x$} = \emph{qualified $x$}, {n/e/p} = \emph{supported for (n)odes, (e)dges, and (p)roperties},

   {o/c} = \emph{(o)pen and (c)losed}, {m/o} = \emph{(m)andatory and (o)ptional},  
   {f/p/x} = \emph{schema (f)irst, (p)artial, and fle(x)ible}, 
   {oC} = \emph{openCypher},
% \end{minipage}
% \end{adjustbox}
\end{table*}

\subsubsection*{Conceptual data models} ER-based data models tend to be agnostic with respect to attribute types, since these may depend on the back-end for which the data model is designed.
%exact types of the attributes, since this may depend on the back-end for which we are designing the data model.
Most support inheritance hierarchies and, in that way, can model union and intersection types.
Entity types can be modelled as abstract types by indicating that their entities must belong to at least one of their subtypes.
Since the final goal is to design a relational schema, which is closed, none of them support open types.
Most ER-based models allow attributes to be composed and/or multi-valued, and so can model complex nested values.
A surprising restriction is that most conceptual data models only allow a single key and require it to be a single attribute.
A notable exception is ORM2, which can support any number of composed keys. 
Schema validation is not applicable in this context, as there is no notion of a schema-independent instance.
%As far as validation goes, this not an applicable notion in this context. This is because most schema formalisms here are schema-full and mapped to a relational schema, and so there is no notion of an instance independent of a schema. 

\subsubsection*{RDF formalisms}
RDF-based formalisms inherit XML datatypes with some limitations (PDT). 
Both SHACL and ShEx are based on a kind of \emph{open} semantics in which the closeness of a constraint needs to be specified with a keyword \emph{close} (OCT). The element properties are expressible only over the nodes, except the recent proposal of RDF-star, that extends RDF exactly with properties over edges (EP). SHACL and ShEx are missing explicit support for key-like constraints (KC), but allow for cardinality constraints (CC), to which SHACL applies set and ShEx bag semantics. The complexity of validation (TV) of RDF-based formalisms is a well-researched topic. While it is not tractable in general for the most expressive cases, practically useful fragments do have this property.

% \textcolor{purple}{Discuss RDF rows here}

\subsubsection*{Tree-structured data}
% Discuss DTDs RELAX NG and XML Schema
DTDs support union types (UIT) in content type expressions by allowing disjunction, despite it not being applicable to the names of XML element types.
%but such disjunction is not possible for the name of an XML element type.
This is, however, possible in both RELAX NG and XML Schema, where we have full union types, although both require the regular expressions that describe the content to be deterministic. 
DTDs do not support inheritance (TH) or abstract types (AT), although element types can be embedded in the content of other element types.
However, RELAX NG and XML Schema have explicit support for these features.
In DTDs, the content of an element can be made open by using \texttt{PCDATA}, which allows any XML fragment as content, so (OCT) is more or less supported.
In RELAX NG and XML Schema, it is possible to have an even more sophisticated mix of open and closed parts within a content type.
Attributes of XML elements can be declared as required and are optional by default.
For some basic attribute types there are multi-valued variants that can be assigned, but, otherwise, there are no complex nested attribute types.
Here again, RELAX NG and XML Schema offer a set of type constructors that enable users to build new and more complex attribute types.
Concerning (KC), both in DTDs and RELAX NG, an attribute can be defined as key, by giving it the type \texttt{ID}, but there is no way to declare an arbitrary set of attributes as key.
XML Schema, on the other hand, has a very elaborate notion of key constraint that can define a complex key for elements where this key consists of several values, is not restricted to attributes of the element, and is applicable only to a specified set of elements.
DTDs support mandatory participation (MP), since they can require that the content of a certain element be nonempty or that a certain attribute reference some document element. 
This is the case in XML Schema, which also supports foreign keys, and RELAX NG. DTDs offer some support for (IS), in the sense that they are regarded as included in the XML document, and so are accessible by software processing the document.
For both XML Schema and RELAX NG there is an XML syntax, and so they allow full introspection.
For DTDs, XML Schema, and RELAX NG, it holds that XML documents can be created without a schema, but it is also possible to restrict updates to only allow those which maintain the document's conformance to the schema; thus, for (SFPX), both schema-first and schema-flexible are supported. Finally, the BonXai language~\cite{DBLP:journals/pvldb/MartensNNS12,DBLP:journals/tods/MartensNNS17} is a research prototype that can define all schemas expressible in XML Schema, yet uses a DTD-like philosophy for specifying its validation rules. It supports XML Schema complex types and constraints, but no type inheritance.

% discuss JSON Schema
JSON Schema supports unions and intersections of open types (UIT), through the \texttt{anyOf} and, respectively, \texttt{allOf} keywords. 
The latter also supports building type hierarchies (TH), but only for open types. 
Whether types are instantiated depends on the choice of the root object, and so there is arguably support for abstract types (AT).
One can control whether types are open or closed by setting \texttt{additionalProperties} to \texttt{True} or \texttt{False}. 
Fields are optional by default, but can be marked as mandatory through the \texttt{required} construct. 
Range constraints (RC) are limited to numerical values and strings. 
Field values support (CPT), as they can encode nested objects and arrays. 
Key constraints (KC) are not supported and cardinality constraints (CC) specify bounds for arrays.  Finally, as the schema is itself a JSON object, it allows introspection (IS).

%\textcolor{purple}{TO BE DISCUSSED: JSON Schema \\
%Some notes:
%\begin{itemize}
%  \item Union types: JSON Schema has \texttt{anyOf} for union types (and \texttt{oneOf}) and \texttt{allOf} (but this only works for open types and cannot be used to extend closed types). 
%  \item Type hierarchies can be done with \texttt{allOf} but only for open types.
%  \item There is no notion of abstract type, but we pick what is instantiated by picking ``the root'' and can accomplish in that way the same effect.
%  \item There is control over types being open and closed with the property \texttt{additionalProperties} set to True or False.
%  \item Fields are optional by default, but can be marked as mandatory by putting them in the list of the field \texttt{required}.
%  \item All fields are in some sense complex values here (objects and arrays nested).
%  \item There are no key constraints as normally understood in database modelling (e.g., declare that in a certain array of objects a certain property identifies each object in the array).
%  \item There are limited cardinality constraints: bounds (lower and upper) of arrays can be specified.
%  \item Schema is itself JSON, and so there is introspection.
%\end{itemize}
%}

\subsubsection*{Existing graph technologies}

We now discuss the extent to which type, constraints, and schema features are supported in several state-of-the-art graph schema languages and systems.

\emph{Type features}, such as union and intersection types (UIT), type hierarchies (TH), and abstract types (AT), are supported by GraphQL, and JSON Schema. 
These capabilities are also found in SQL-based systems, such as OrientDB/SQL, and in systems able to leverage JSON Schema, such as ArangoDB and AgensGraph.
The strongly-typed TypeDB database has a rich type system that also offers subtyping for entities, relations, and attributes/properties. 
While other considered systems cannot directly handle TH, some can emulate it through multi-labels by adding all the intended parent types of a node as its additional labels. 
Nevertheless, this is problematic for validation, as one cannot ensure that all subtypes have been assigned the correct label. 
Note that most examined technologies only implement closed types (OCT), except ArangoDB and OrientDB, in which also open types are possible, and Neo4j, which considers types open and, hence, extensible, by default.

Nodes and edges can be enriched with element properties (EP) in all surveyed graph technologies, and in AgensGraph these can be defined using JSON objects. 
Such properties are optional by default in AgensGraph, ArangoDB, DataStax, JanusGraph, Neo4j, and Sparksee, though users can define mandatory constraints to enforce them being non-nullable. 
In Nebula Graph, users can specify, when designing their schema, whether null-valued attributes are allowed, while in TypeDB these are not supported. 
Finally, the (MOP) feature is present in SQL-based technologies. 
Most reviewed graph languages and system also allow for (CPT), although specific restrictions sometimes apply. 
For example, in Neo4j, complex property values can only be homogeneous lists of simple types and byte arrays, despite the latter not being first-class Cypher data types. 
AgensGraph draws its support for (CPT) from openCypher and JSON, while in Sparksee, multi-valued properties can only be defined using array attributes, using all but the \texttt{String} and \texttt{Text} data types. In addition, range constraints (RC) can be specified for any data type, in SQL-based technologies, and for numerical values and strings, in systems that build on JSON Schema or that provide regular expressions, such as TypeDB.

Regarding \emph{constraint features}, we remark that key constraints are available in all reviewed graph schema languages except GQL. 
At a system level, AgensGraph, Neo4j, and Sparksee support node uniqueness constraints, disallowing the same property values from appearing in more than one node of a given label or type, while ArangoDB enables specifying \texttt{uniqueItems} for arrays, thanks to JSON Schema. 
Some technologies, such as DataStax, Oracle/PGQL, and TigerGraph, offer primary keys for nodes, which enforce property values to be unique, mandatory, and single-valued. 
TigerGraph GSQL also supports the notion of a discriminator, which is an attribute or set of attributes that can be used to uniquely identify an edge, when multiple instances of a given type exist between a pair of vertices. 
Finally, the considered SQL-based systems can rely on SQL's mechanism for defining unique key constraints for tables. These systems also feature mandatory participation (MP) and uniqueness constraints.

More general forms of cardinality constraints (CC) are only  provided by a few systems. Among these, JanusGraph allows declaring edge label multiplicity: the \texttt{MULTI} and \texttt{SIMPLE} keywords can specify whether multiple edges or at most one can be defined between 
any node pair; \texttt{MANY2ONE} and \texttt{ONE2MANY} respectively allow at most one outgoing/incoming edge, without constraining the number of incoming/outgoing ones. The system also provides property key cardinalities, i.e., declaring whether one (\texttt{SINGLE}), an arbitrary number (\texttt{LIST}), or multiple, non-duplicate values (\texttt{SET}) can be associated with a node key. Other examples include ArangoDB, leveraging JSON Schema's \texttt{minProperties}/\texttt{maxProperties} keywords to restrict the number of object properties, and 
TypeDB, providing high-level CCs at the type level that require relationships to have at least one role that specifies their nature. TypeDB is also the only system that handles binary-relation constraints (BRC), such as symmetry and transitivity, by expressing them via inference rules. In TigerGraph, the support for (BRC) is limited to declaring reverse edge types.

Tractable validation (TV) is a \emph{schema feature} supported by systems that leverage JSON Schema and SQL. In JanusGraph and Sparksee, the schema is defined through their specific APIs and there is no formal account of their schema validation mechanisms. 
Concerning introspection (IS), all reviewed systems support it either directly, through the query language itself, or indirectly, via a management API (like in JanusGraph and Sparksee). 
Finally, the only reviewed system that natively supports (SPFX) is OrientDB/SQL, which has schema-first, schema-less, and explicit schema-hybrid modes. Partial conformance is possible in all systems that are not exclusively schema-first or that do not have native schema mechanisms, like ArangoDB, which relies on JSON Schema.

\subsubsection*{\pgschema} 
Like SQL/PGQ and GQL, \pgschema views a set of node and edge types as the core of a graph database schema. The support for type features is essentially complete, as discussed in Section~\ref{sec:proposal}, except that (CPT) and (RC), as well as (UIT) and (TH) for properties, are delegated to the property type system. While concrete property types have been used in examples, \pgschema deliberately leaves the choice of the property type system open, which allows it to function as an embedded language in both GQL and SQL/PGQ, offering suitable property type features. In Section~\ref{ssec:extensions} we discuss how some of these features could be supported directly in \pgschema.

In terms of constraint features, \pgschema is also quite comprehensive. As discussed in Section~\ref{ssec:constraints}, it supports not only (KC) and (MP), but also denial constraints and uniqueness constraints.
In Section~\ref{ssec:extensions} we discuss an extension to support general cardinality constraints (CC) that fits well with the support for (KC) and (MP).

Important design principles for \pgschema were to preserve the spirit of schemas in SQL/PGQ and graph types in GQL, keeping node and edge types locally verifiable, while at the same time offering a powerful mechanism to express constraints. This is why \pgkeys are clearly separated from \pgtypes, and constraints can refer to types, but types cannot refer to constraints. 
% <<<<<<< Updated upstream
% This is in contrast with other approaches, such as SHACL \textcolor{red}{and the very similar ProGS \cite{Seifer21}}, where there is no such separation and a small change in one element may affect the types of many distant elements.  
% =======
%
%
% old paragraph 
% This is in contrast with other approaches, such as SHACL, where there is no such separation and a small change in one element may affect the types of many distant elements.  
%
% new longer paragraph 
%
% \textcolor{blue}{
This is in contrast with other approaches, such as SHACL and ShEx where 
new types (there also called shapes or labels) can be assigned to neighboring nodes during the verification process. As a result, new types are propagated throughout the graphs and more nodes and types need to be checked, i.e., a node type may affect types of some distant nodes. Such an approach is in particular problematic when types have circular definitions and this issue has been left open in the  SHACL standard. Recently, there have been several proposals that addresses this issue for SHACL~\cite{DBLP:conf/semweb/CormanRS18,DBLP:conf/www/AndreselCORSS20} and ShEx~\cite{Staworko2015complexity}, borrowing the ideas from logic programming. The same approach is taken in ProGS \cite{Seifer21} that introduces SHACL-like constraints for property graphs.
% }

Because types are locally verifiable, \pgschema has tractable validation (TV), as long as each constraint alone is tractable, which is the case for key, participation, and cardinality constraints.
\pgschema fully supports (SFPX), as it allows defining strict schemas (schema first), loose schemas that only enforce constraints on typed elements (partial schema), and schemas that allow every graph (flexible schema);
note that it is possible to generate a descriptive schema, as required (see Section~\ref{ssec:validation}).
Finally, owing to locally verifiable types and the design of \pgkeys, if a graph only partially conforms to a given schema (strict or loose), this can be easily explained to users by indicating which elements are typed and which  satisfy the constraints, thus supporting meaningful partial validation.

Basic graph types can be naturally represented as property graphs, as shown in Figure~\ref{fig-uc2-diagram}.  However, there is currently no commonly agreed-upon way of reflecting the powerful mechanism of type combinations in \pgtypes; hence,  introspection (IS) is not supported.
How such a representation can capture all features  and yet remain intuitive, is an issue for future research.

%There is currently no commonly agreed upon definition within \pgschema of how schemas are represented as property graphs, and hence it does not fully support introspection (IS). How this representation can capture all features and yet remain intuitive, as in the diagram of Figure~\ref{fig-uc2-diagram}, is an issue for future research.

\subsubsection*{Conclusion}
The table shows that there is quite some variety in which features are supported and also that no formalism or system covers all of them. 
This likely reflects that they tend to target different sets of use cases.
Likewise, \pgschema does not attempt to cover all features, but it does aim to provide a foundation that could be extended to do so.

\subsection{Possible Extensions of \pgschema}
\label{ssec:extensions}

% \textcolor{purple}{Intended topics: 
% \begin{itemize}
%     \item expressive content type language
%     \item complex properties
%     \item more expressive constraint language (cardinality constraints, denial constraints)
% \end{itemize}
% }

%\subsubsection{Expressive content type language}
%\textcolor{purple}{Under construction}
Let us now discuss briefly how some of the currently unsupported features from Table~\ref{tab:featureTable} could be integrated into \pgschema.
%In this subsection we take the features from Table~\ref{tab:featureTable} missing in \pgschema, and discuss how the latter can be extended in order to support them.

\subsubsection*{Range constraints.} Some schema languages allow for range constraints (RC). The syntax of \pgschema can be thus extended, specifying restrictions on acceptable values for properties. For instance, the following example defines a node type \inlinecode{Book}, with properties \inlinecode{title} (a string with maximum 100 characters), \inlinecode{genre} (an enumeration), and \inlinecode{isbn} (a string conforming to a regular expression):
%\begin{lstlisting}
%NODE TYPE $Person (:Person CLOSED) {
%    givenName : STRING MinLength 1 ,
%    familyName : STRING MinLength 2 MaxLength 20 ,
%    birthday : DATE ((19|20)\d{2}(0[1-9]|1[0-2])(0[1-9]|[12]\d|3[01]))
%    CLOSED
%}
%\end{lstlisting}
\begin{lstlisting}
  (bookType: Book {
    title STRING(100),
    genre ENUM("Prose", "Poetry", "Dramatic"),
    isbn STRING ^(?=(?:\D*\d){10}(?:(?:\D*\d){3})?$)[\d-]+$})
\end{lstlisting}
The restrictions allowed in the example above can be based on XSD facets \cite{Peterson2012WXS}, with additional features, such as enumerations, implemented similarly. %Features such as enumerations can be similarly implemented.

\subsubsection*{Complex datatypes.}
We have only included primitive datatypes in \pgschema. Looking at the CPT column in Table \ref{tab:featureTable} we see that complex property values are widely supported in other formalisms. Our syntax can easily be extended to support collections. For instance, if we wish to specify that a \inlinecode{name} is an array of strings, we could write \inlinecode{name STRING ARRAY \{1,2\}}. The (optional) annotation in curly braces specifies the minimum  and the maximum number  of elements in the array.
%The array can optionally have a minimum (the first value in curly brackets) and maximum number (the second value in curly brackets) of elements.

% \domagoj{Not sure how to proceed with the example below. I would state that this can be simulated with different types already (e.g. abstract ones, where one has name, the other the family/given name option). However, the ability to conform to different types is valuable, so I would allow this with an explicit OneOf command. For instance \inlinecode{name: OneOf\{STRING, INT\}}, or something similar.}

\subsubsection*{Intersections and unions for content types.}
In Section \ref{sec:proposal}, we assumed that union and intersection can be used in element types. 
Such combinations can also be introduced into properties (see UIT in Table~\ref{tab:featureTable}). 
We can for example allow the \emph{union} (\inlinecode{|}) and \emph{intersection} (\inlinecode{\&}) operators from label expressions also between property types and between content types.
% \dominik{The first union is exclusive or, do we really want to use the same symbol for or and xor?}
% \jan{It's in all cases the inclusive union, which in this case just happens to be exclusive because of the types is combines.}
% To implement combinations of properties we now redefine a comma as \emph{EachOf} operator, then add a caret sign as \emph{OneOf} operator. The \emph{OneOf} operator can be used also in datatypes (e.g. \inlinecode{INT \^{} FLOAT}). 
The following example allows the \inlinecode{name} to be broken down into a \inlinecode{givenName} and a \inlinecode{familyName}:
% \dominik{Maybe I don't understand a new syntax, but IMO there is sth wrong with brackets in properties. Curly brackets are for definition of property (and below I see round brackets). Round brackets are for grouping (and below I see curly brackets)} 
% \jan{The content expression has the form \texttt{(\{..\}) | (\{..\}) \{..\}}. Here the \texttt{..} are lists of property declarations. So indeed the round brackets are used for grouping content definitions (or rather determine the order in which the boolean operators are applied that combine them) and the curly braces indeed identify content definitions.}
\begin{lstlisting}
(personType: Person 
  ( {name STRING} | {givenName STRING, familyName STRING} )
  & {height (INT | FLOAT)})
\end{lstlisting}
% \begin{lstlisting}
% NODE TYPE #Person (:Person) {
%   name:  STRING ^
%   (givenName :  STRING ,
%   familyName : STRING) ,
%   height : INT ^ FLOAT 
% }
% \end{lstlisting}
Note that we use round brackets to group content definitions. 
% Choices of properties can be simulated by abstract types.

\subsubsection*{Advanced cardinalities.}
The notation of \pgkeys can be readily extended to specify cardinality constraints by taking the general form
% \begin{lstlisting}[mathescape=true]
%  FOR $p(x)$ <qualifier> $q(x,\bar y)$(*\comma*)
%\end{lstlisting}
\lstinline[mathescape=true]{FOR $p(x)$ <qualifier> $q(x,\bar y)$},
and allowing for \lstinline{<qualifier>} an expression of the form \lstinline{COUNT <lower bound>?..<upper bound>? OF}, expressing that the number of distinct results returned by $q(x, \bar y)$ must be within that range.
If the upper bound and lower bound are identical, we allow the short-hand \lstinline{COUNT <bound> OF}.

The constraint stating that each department has at least two employees working for it, could be written as
\begin{lstlisting}[mathescape=true]
  FOR (d: Department) 
    COUNT 2.. OF e WITHIN (e: Employee)-[:worksIn]->(d)(*\period*)
\end{lstlisting}
And if employees can work on at most 3 projects, this could be written as
\begin{lstlisting}[mathescape=true]
  FOR (e: Employee) 
    COUNT 0..3 OF p WITHIN (e)-[: worksOn]->(p: Project)(*\period*)
\end{lstlisting}
This notation also allows us to express disjointness and denial constraints without using negation in patterns. 
For example, if reptiles cannot be amphibians, we can write this as
\begin{lstlisting}[mathescape=true]
  FOR (a: Amphibian) 
    COUNT 0 OF (a: Reptile)(*\period*)
\end{lstlisting}
As discussed in \cite{Threshold:2022}, such constraints are relevant for many practical use cases and can be efficiently evaluated.

% \textcolor{purple}{TO BE WRITTEN (see some examples in LaTeX comment lines for suggested syntax based on PG-Keys)}

% \textcolor{gray}{
% \begin{description}
% \item[*] 0 or more,
% \item[+] 1 or more,
% \item[?] 0 or 1,
% \item{\{m\}} exactly $m$ repetitions,
% \item[\{m,n\}] between $m$ and $n$ repetitions,
% \item[\{m,\}] $m$ or more repetitions.
% \end{description}
% }

% Departments must have at least 2 employees.
%
% FOR (d:Department) 
% COUNT {2,} OF e 
% WITHIN (e:Employee)-[:worksIn]->(d)

% Employees must not work on more than 3 projects.
%
% FOR (e:Employee) 
% COUNT {,3} OF p
% WITHIN (e)-[:worksOn]->(p:Project)

% zoom arraignment
%
% FOR (x)-[]->() 
% COUNT {1,10} OF (y, z) 
% WITHIN (x)-[y]->(z) 

%FOR ($Amphibian & $Reptile) MANDATORY @false
%FOR ($Amphibian) MANDATORY (!$Reptile)

% cardinality constraints, denial constraints

% negation in label sets?

%%% Local Variables:
%%% mode: latex
%%% TeX-master: "main"
%%% End:
\section{Summary and Looking ahead}
\label{sec:conclusions}
\pgschema is the first unifying schema language for property graphs, which serves as a recommendation for future versions of GQL. 
This work is the result of academia and industry collaborating to bridge gaps and accelerate standardization efforts that benefit both communities at large. 

\subsubsection*{Summary}
\pgschema is a schema language that caters to basic needs such as defining node and edge types, as well as advanced scenarios such as expressing complex type hierarchies and integrity constraints.
It has been designed to support both descriptive and prescriptive roles, with a focus on enabling agile evolution, flexible validation, and usability. 
The language comes with an ASCII-art, yet formal, syntax and well-defined semantics. 
The core of \pgschema centers around the rich \pgtypes type system, with desirable features such as compositionality, abstract types, type hierarchies, and multi-inheritance, as well as around \pgkeys, which allows the expression of complex key and participation constraints. 
The language thus supports a wide-range of capabilities, largely absent from the state-of-the-art schema languages and systems we have reviewed. 
Finally, \pgschema is easily extensible with further features, such as range constraints, complex data types, content type combinators, and advanced cardinalities. 

\subsubsection*{Looking Ahead}
In addition to the impact on standardization efforts, this is an opportunity for graph database vendors to increase functionality and support current and future customer demands. 
Our work also provides a basis for future research by the academic community. 
Finally, this successful high-impact academia-industry collaboration model is one we hope will be replicated by other communities at large, in data management and beyond.

%%% Local Variables:
%%% mode: latex
%%% TeX-master: "main"
%%% End:

\begin{acks}
We thank all LDBC PGSWG members for the discussions around property graph schemas. Renzo Angles was supported by ANID Millennium Science Initiative Program, Code ICN17\_002 and ANID FONDECYT Chile through grant 1221727. Angela Bonifati and Leonid Libkin were supported by ANR-21-CE48-0015 VeriGraph. Leonid Libkin was supported by a Leverhulme Trust Research fellowship and EPSRC grant S003800. Wim Martens was
supported by ANR project EQUUS ANR-19-CE48-0019; funded by the Deutsche
Forschungsgemeinschaft (DFG, German Research Foundation) -- project number
431183758. Filip Murlak was supported by NCN grant 2018/30/E/ST6/00042. Domagoj Vrgo\v{c} was supported by ANID Millennium Science Initiative Program, Code ICN17\_002 and ANID Fondecyt Regular project 1221799. 
For the purposes of open access, the authors have applied a CC BY public copyright licence to any Author Accepted Manuscript version arising from this submission.
\end{acks}

\bibliographystyle{ACM-Reference-Format}
\bibliography{bibliography}

%%% -*-BibTeX-*-
%%% Do NOT edit. File created by BibTeX with style
%%% ACM-Reference-Format-Journals [18-Jan-2012].

\begin{thebibliography}{65}

%%% ====================================================================
%%% NOTE TO THE USER: you can override these defaults by providing
%%% customized versions of any of these macros before the \bibliography
%%% command.  Each of them MUST provide its own final punctuation,
%%% except for \shownote{}, \showDOI{}, and \showURL{}.  The latter two
%%% do not use final punctuation, in order to avoid confusing it with
%%% the Web address.
%%%
%%% To suppress output of a particular field, define its macro to expand
%%% to an empty string, or better, \unskip, like this:
%%%
%%% \newcommand{\showDOI}[1]{\unskip}   % LaTeX syntax
%%%
%%% \def \showDOI #1{\unskip}           % plain TeX syntax
%%%
%%% ====================================================================

\ifx \showCODEN    \undefined \def \showCODEN     #1{\unskip}     \fi
\ifx \showDOI      \undefined \def \showDOI       #1{#1}\fi
\ifx \showISBNx    \undefined \def \showISBNx     #1{\unskip}     \fi
\ifx \showISBNxiii \undefined \def \showISBNxiii  #1{\unskip}     \fi
\ifx \showISSN     \undefined \def \showISSN      #1{\unskip}     \fi
\ifx \showLCCN     \undefined \def \showLCCN      #1{\unskip}     \fi
\ifx \shownote     \undefined \def \shownote      #1{#1}          \fi
\ifx \showarticletitle \undefined \def \showarticletitle #1{#1}   \fi
\ifx \showURL      \undefined \def \showURL       {\relax}        \fi
% The following commands are used for tagged output and should be
% invisible to TeX
\providecommand\bibfield[2]{#2}
\providecommand\bibinfo[2]{#2}
\providecommand\natexlab[1]{#1}
\providecommand\showeprint[2][]{arXiv:#2}

\bibitem[\protect\citeauthoryear{39075}{39075}{2023}]%
        {ISO39075}
\bibfield{author}{\bibinfo{person}{ISO/IEC 39075}.}
  \bibinfo{year}{2023}\natexlab{}.
\newblock \bibinfo{booktitle}{\emph{Information technology — Database
  languages — GQL}}.
\newblock \bibinfo{type}{Standard}. \bibinfo{institution}{International
  Organization for Standardization}, \bibinfo{address}{Geneva, CH}.
\newblock


\bibitem[\protect\citeauthoryear{9075-16}{9075-16}{2022}]%
        {ISO9075}
\bibfield{author}{\bibinfo{person}{ISO/IEC 9075-16}.}
  \bibinfo{year}{2022}\natexlab{}.
\newblock \bibinfo{booktitle}{\emph{Information technology — Database
  languages SQL — Part 16: Property Graph Queries (SQL/PGQ)}}.
\newblock \bibinfo{type}{Standard}. \bibinfo{institution}{International
  Organization for Standardization}, \bibinfo{address}{Geneva, CH}.
\newblock


\bibitem[\protect\citeauthoryear{AgensGraph}{AgensGraph}{2022}]%
        {AgensGDB}
\bibfield{author}{\bibinfo{person}{AgensGraph}.}
  \bibinfo{year}{2022}\natexlab{}.
\newblock \bibinfo{title}{AgensGraph}.
\newblock \bibinfo{howpublished}{\url{https://bitnine.net/agensgraph} (visited:
  2022-11)}.
\newblock


\bibitem[\protect\citeauthoryear{Andresel, Corman, Ortiz, Reutter, Savkovic,
  and Simkus}{Andresel et~al\mbox{.}}{2020}]%
        {DBLP:conf/www/AndreselCORSS20}
\bibfield{author}{\bibinfo{person}{Medina Andresel}, \bibinfo{person}{Julien
  Corman}, \bibinfo{person}{Magdalena Ortiz}, \bibinfo{person}{Juan~L.
  Reutter}, \bibinfo{person}{Ognjen Savkovic}, {and} \bibinfo{person}{Mantas
  Simkus}.} \bibinfo{year}{2020}\natexlab{}.
\newblock \showarticletitle{Stable Model Semantics for Recursive {SHACL}}. In
  \bibinfo{booktitle}{\emph{{WWW} '20: The Web Conference 2020, Taipei, Taiwan,
  April 20-24, 2020}}, \bibfield{editor}{\bibinfo{person}{Yennun Huang},
  \bibinfo{person}{Irwin King}, \bibinfo{person}{Tie{-}Yan Liu}, {and}
  \bibinfo{person}{Maarten van Steen}} (Eds.). \bibinfo{publisher}{{ACM} /
  {IW3C2}}, \bibinfo{pages}{1570--1580}.
\newblock
\urldef\tempurl%
\url{https://doi.org/10.1145/3366423.3380229}
\showDOI{\tempurl}


\bibitem[\protect\citeauthoryear{Angles, Arenas, Barcel{\'o}, Boncz, Fletcher,
  Gutierrez, Lindaaker, Paradies, Plantikow, Sequeda, van Rest, and
  Voigt}{Angles et~al\mbox{.}}{2018}]%
        {gcore}
\bibfield{author}{\bibinfo{person}{Renzo Angles}, \bibinfo{person}{Marcelo
  Arenas}, \bibinfo{person}{Pablo Barcel{\'o}}, \bibinfo{person}{Peter~A.
  Boncz}, \bibinfo{person}{George H.~L. Fletcher}, \bibinfo{person}{Claudio
  Gutierrez}, \bibinfo{person}{Tobias Lindaaker}, \bibinfo{person}{Marcus
  Paradies}, \bibinfo{person}{Stefan Plantikow}, \bibinfo{person}{Juan~F.
  Sequeda}, \bibinfo{person}{Oskar van Rest}, {and} \bibinfo{person}{Hannes
  Voigt}.} \bibinfo{year}{2018}\natexlab{}.
\newblock \showarticletitle{{G-CORE:} {A} Core for Future Graph Query
  Languages}. In \bibinfo{booktitle}{\emph{{SIGMOD} Conference}}.
  \bibinfo{publisher}{{ACM}}, \bibinfo{pages}{1421--1432}.
\newblock


\bibitem[\protect\citeauthoryear{Angles, Bonifati, Dumbrava, Fletcher, Hare,
  Hidders, Lee, Li, Libkin, Martens, Murlak, Perryman, Savkovic, Schmidt,
  Sequeda, Staworko, and Tomaszuk}{Angles et~al\mbox{.}}{2021}]%
        {PGkeys}
\bibfield{author}{\bibinfo{person}{Renzo Angles}, \bibinfo{person}{Angela
  Bonifati}, \bibinfo{person}{Stefania Dumbrava}, \bibinfo{person}{George
  Fletcher}, \bibinfo{person}{Keith~W. Hare}, \bibinfo{person}{Jan Hidders},
  \bibinfo{person}{Victor~E. Lee}, \bibinfo{person}{Bei Li},
  \bibinfo{person}{Leonid Libkin}, \bibinfo{person}{Wim Martens},
  \bibinfo{person}{Filip Murlak}, \bibinfo{person}{Josh Perryman},
  \bibinfo{person}{Ognjen Savkovic}, \bibinfo{person}{Michael Schmidt},
  \bibinfo{person}{Juan~F. Sequeda}, \bibinfo{person}{Slawek Staworko}, {and}
  \bibinfo{person}{Dominik Tomaszuk}.} \bibinfo{year}{2021}\natexlab{}.
\newblock \showarticletitle{PG-Keys: Keys for Property Graphs}. In
  \bibinfo{booktitle}{\emph{International Conference on Management of Data
  (SIGMOD)}}. \bibinfo{publisher}{{ACM}}, \bibinfo{pages}{2423--2436}.
\newblock


\bibitem[\protect\citeauthoryear{ArangoDB}{ArangoDB}{2022}]%
        {arango:website}
\bibfield{author}{\bibinfo{person}{ArangoDB}.} \bibinfo{year}{2022}\natexlab{}.
\newblock \bibinfo{title}{ArangoDB}.
\newblock \bibinfo{howpublished}{\url{https://www.arangodb.com/} (visited:
  2022-11)}.
\newblock


\bibitem[\protect\citeauthoryear{Baker and Prud'hommeaux}{Baker and
  Prud'hommeaux}{2019}]%
        {Baker2019}
\bibfield{author}{\bibinfo{person}{Thomas Baker} {and} \bibinfo{person}{Eric
  Prud'hommeaux}.} \bibinfo{year}{2019}\natexlab{}.
\newblock \bibinfo{booktitle}{\emph{Shape Expressions ({ShEx}) 2.1 Primer}}.
\newblock \bibinfo{type}{{W3C} Community Group Final Report}.
  \bibinfo{institution}{W3C}.
\newblock
\newblock
\shownote{\url{https://shex.io/shex-primer/index.html}.}


\bibitem[\protect\citeauthoryear{Bonifati, Dumbrava, Fletcher, Hidders, Hofer,
  Martens, Murlak, Shinavier, Staworko, and Tomaszuk}{Bonifati
  et~al\mbox{.}}{2022a}]%
        {Threshold:2022}
\bibfield{author}{\bibinfo{person}{Angela Bonifati}, \bibinfo{person}{Stefania
  Dumbrava}, \bibinfo{person}{George Fletcher}, \bibinfo{person}{Jan Hidders},
  \bibinfo{person}{Matthias Hofer}, \bibinfo{person}{Wim Martens},
  \bibinfo{person}{Filip Murlak}, \bibinfo{person}{Joshua Shinavier},
  \bibinfo{person}{S\l{}awek Staworko}, {and} \bibinfo{person}{Dominik
  Tomaszuk}.} \bibinfo{year}{2022}\natexlab{a}.
\newblock \showarticletitle{Threshold Queries in Theory and in the Wild}.
\newblock \bibinfo{journal}{\emph{Proc. VLDB Endow.}} \bibinfo{volume}{15},
  \bibinfo{number}{5} (\bibinfo{date}{may} \bibinfo{year}{2022}),
  \bibinfo{pages}{1105–1118}.
\newblock
\showISSN{2150-8097}
\urldef\tempurl%
\url{https://doi.org/10.14778/3510397.3510407}
\showDOI{\tempurl}


\bibitem[\protect\citeauthoryear{Bonifati, Dumbrava, Fletcher, Hidders, Li,
  Libkin, Martens, Murlak, Plantikow, Savković, Sequeda, Staworko, Tomaszuk,
  Voigt, Vrgoč, and Wu}{Bonifati et~al\mbox{.}}{2022b}]%
        {grammar}
\bibfield{author}{\bibinfo{person}{Angela Bonifati}, \bibinfo{person}{Stefania
  Dumbrava}, \bibinfo{person}{George Fletcher}, \bibinfo{person}{Jan Hidders},
  \bibinfo{person}{Bei Li}, \bibinfo{person}{Leonid Libkin},
  \bibinfo{person}{Wim Martens}, \bibinfo{person}{Filip Murlak},
  \bibinfo{person}{Stefan Plantikow}, \bibinfo{person}{Ognjen Savković},
  \bibinfo{person}{Juan Sequeda}, \bibinfo{person}{Sławek Staworko},
  \bibinfo{person}{Dominik Tomaszuk}, \bibinfo{person}{Hannes Voigt},
  \bibinfo{person}{Domagoj Vrgoč}, {and} \bibinfo{person}{Mingxi Wu}.}
  \bibinfo{year}{2022}\natexlab{b}.
\newblock \showarticletitle{domel/pgschema: PG-Schema Grammar 0.3}.
\newblock  (\bibinfo{date}{Nov} \bibinfo{year}{2022}).
\newblock
\urldef\tempurl%
\url{https://doi.org/10.5281/zenodo.7362078}
\showDOI{\tempurl}
\newblock
\shownote{https://zenodo.org/record/7362078.}


\bibitem[\protect\citeauthoryear{Bonifati, Dumbrava, Fletcher, Hidders, Li,
  Libkin, Martens, Murlak, Plantikow, Savković, Sequeda, Staworko, Tomaszuk,
  Voigt, Vrgoč, and Wu}{Bonifati et~al\mbox{.}}{2022c}]%
        {pgsarxiv}
\bibfield{author}{\bibinfo{person}{Angela Bonifati}, \bibinfo{person}{Stefania
  Dumbrava}, \bibinfo{person}{George Fletcher}, \bibinfo{person}{Jan Hidders},
  \bibinfo{person}{Bei Li}, \bibinfo{person}{Leonid Libkin},
  \bibinfo{person}{Wim Martens}, \bibinfo{person}{Filip Murlak},
  \bibinfo{person}{Stefan Plantikow}, \bibinfo{person}{Ognjen Savković},
  \bibinfo{person}{Juan Sequeda}, \bibinfo{person}{Sławek Staworko},
  \bibinfo{person}{Dominik Tomaszuk}, \bibinfo{person}{Hannes Voigt},
  \bibinfo{person}{Domagoj Vrgoč}, {and} \bibinfo{person}{Mingxi Wu}.}
  \bibinfo{year}{2022}\natexlab{c}.
\newblock \bibinfo{title}{PG-Schema: Schemas for Property Graphs}.
\newblock
\newblock
\urldef\tempurl%
\url{https://doi.org/10.48550/arXiv.2211.10962}
\showDOI{\tempurl}


\bibitem[\protect\citeauthoryear{Bonifati, Dumbrava, Martinez, Ghasemi,
  Jaffr{\'{e}}, Luton, and Pickles}{Bonifati et~al\mbox{.}}{2022d}]%
        {BonifatiDMGJLP22}
\bibfield{author}{\bibinfo{person}{Angela Bonifati},
  \bibinfo{person}{Stefania{-}Gabriela Dumbrava}, \bibinfo{person}{Emile
  Martinez}, \bibinfo{person}{Fatemeh Ghasemi}, \bibinfo{person}{Malo
  Jaffr{\'{e}}}, \bibinfo{person}{Pacome Luton}, {and} \bibinfo{person}{Thomas
  Pickles}.} \bibinfo{year}{2022}\natexlab{d}.
\newblock \showarticletitle{DiscoPG: Property Graph Schema Discovery and
  Exploration}.
\newblock \bibinfo{journal}{\emph{Proc. {VLDB} Endow.}} \bibinfo{volume}{15},
  \bibinfo{number}{12} (\bibinfo{year}{2022}), \bibinfo{pages}{3654--3657}.
\newblock
\urldef\tempurl%
\url{https://www.vldb.org/pvldb/vol15/p3654-bonifati.pdf}
\showURL{%
\tempurl}


\bibitem[\protect\citeauthoryear{Bonifati, Furniss, Green, Harmer, Oshurko, and
  Voigt}{Bonifati et~al\mbox{.}}{2019}]%
        {BonifatiFGHOV19}
\bibfield{author}{\bibinfo{person}{Angela Bonifati}, \bibinfo{person}{Peter
  Furniss}, \bibinfo{person}{Alastair Green}, \bibinfo{person}{Russ Harmer},
  \bibinfo{person}{Eugenia Oshurko}, {and} \bibinfo{person}{Hannes Voigt}.}
  \bibinfo{year}{2019}\natexlab{}.
\newblock \showarticletitle{Schema Validation and Evolution for Graph
  Databases}. In \bibinfo{booktitle}{\emph{{ER}}}
  \emph{(\bibinfo{series}{Lecture Notes in Computer Science},
  Vol.~\bibinfo{volume}{11788})}. \bibinfo{publisher}{Springer},
  \bibinfo{pages}{448--456}.
\newblock


\bibitem[\protect\citeauthoryear{Bracha and Cook}{Bracha and Cook}{1990}]%
        {BrachaC90}
\bibfield{author}{\bibinfo{person}{Gilad Bracha} {and}
  \bibinfo{person}{William~R. Cook}.} \bibinfo{year}{1990}\natexlab{}.
\newblock \showarticletitle{Mixin-based Inheritance}. In
  \bibinfo{booktitle}{\emph{Conference on Object-Oriented Programming Systems,
  Languages, and Applications / European Conference on Object-Oriented
  Programming, {OOPSLA/ECOOP} 1990, Ottawa, Canada, October 21-25, 1990,
  Proceedings}}, \bibfield{editor}{\bibinfo{person}{Akinori Yonezawa}} (Ed.).
  \bibinfo{publisher}{{ACM}}, \bibinfo{pages}{303--311}.
\newblock
\urldef\tempurl%
\url{https://doi.org/10.1145/97945.97982}
\showDOI{\tempurl}


\bibitem[\protect\citeauthoryear{Brickley and Guha}{Brickley and Guha}{2014}]%
        {Brickley2014RS}
\bibfield{author}{\bibinfo{person}{Dan Brickley} {and}
  \bibinfo{person}{Ramanathan Guha}.} \bibinfo{year}{2014}\natexlab{}.
\newblock \bibinfo{booktitle}{\emph{{RDF} Schema {1.1}}}.
\newblock \bibinfo{type}{{W3C} Recommendation}. \bibinfo{institution}{W3C}.
\newblock
\newblock
\shownote{\url{https://www.w3.org/TR/2014/REC-rdf-schema-20140225/}.}


\bibitem[\protect\citeauthoryear{Chen}{Chen}{1976}]%
        {Chen1976}
\bibfield{author}{\bibinfo{person}{Peter~P. Chen}.}
  \bibinfo{year}{1976}\natexlab{}.
\newblock \showarticletitle{The Entity-Relationship Model - Toward a Unified
  View of Data}.
\newblock \bibinfo{journal}{\emph{{ACM} Trans. Database Syst.}}
  \bibinfo{volume}{1}, \bibinfo{number}{1} (\bibinfo{year}{1976}),
  \bibinfo{pages}{9--36}.
\newblock


\bibitem[\protect\citeauthoryear{Corman, Reutter, and Savkovic}{Corman
  et~al\mbox{.}}{2018}]%
        {DBLP:conf/semweb/CormanRS18}
\bibfield{author}{\bibinfo{person}{Julien Corman}, \bibinfo{person}{Juan~L.
  Reutter}, {and} \bibinfo{person}{Ognjen Savkovic}.}
  \bibinfo{year}{2018}\natexlab{}.
\newblock \showarticletitle{Semantics and Validation of Recursive {SHACL}}. In
  \bibinfo{booktitle}{\emph{The Semantic Web - {ISWC} 2018 - 17th International
  Semantic Web Conference, Monterey, CA, USA, October 8-12, 2018, Proceedings,
  Part {I}}} \emph{(\bibinfo{series}{Lecture Notes in Computer Science},
  Vol.~\bibinfo{volume}{11136})}. \bibinfo{publisher}{Springer},
  \bibinfo{pages}{318--336}.
\newblock
\urldef\tempurl%
\url{https://doi.org/10.1007/978-3-030-00671-6\_19}
\showDOI{\tempurl}


\bibitem[\protect\citeauthoryear{DataStax}{DataStax}{2022}]%
        {datastax:website}
\bibfield{author}{\bibinfo{person}{DataStax}.} \bibinfo{year}{2022}\natexlab{}.
\newblock \bibinfo{title}{DataStax}.
\newblock \bibinfo{howpublished}{\url{https://www.datastax.com/} (visited:
  2022-11)}.
\newblock


\bibitem[\protect\citeauthoryear{Deutsch, Francis, Green, Hare, Li, Libkin,
  Lindaaker, Marsault, Martens, Michels, Murlak, Plantikow, Selmer, van Rest,
  Voigt, Vrgoc, Wu, and Zemke}{Deutsch et~al\mbox{.}}{2022}]%
        {DeutschFGHLLLMM22}
\bibfield{author}{\bibinfo{person}{Alin Deutsch}, \bibinfo{person}{Nadime
  Francis}, \bibinfo{person}{Alastair Green}, \bibinfo{person}{Keith Hare},
  \bibinfo{person}{Bei Li}, \bibinfo{person}{Leonid Libkin},
  \bibinfo{person}{Tobias Lindaaker}, \bibinfo{person}{Victor Marsault},
  \bibinfo{person}{Wim Martens}, \bibinfo{person}{Jan Michels},
  \bibinfo{person}{Filip Murlak}, \bibinfo{person}{Stefan Plantikow},
  \bibinfo{person}{Petra Selmer}, \bibinfo{person}{Oskar van Rest},
  \bibinfo{person}{Hannes Voigt}, \bibinfo{person}{Domagoj Vrgoc},
  \bibinfo{person}{Mingxi Wu}, {and} \bibinfo{person}{Fred Zemke}.}
  \bibinfo{year}{2022}\natexlab{}.
\newblock \showarticletitle{Graph Pattern Matching in {GQL} and {SQL/PGQ}}. In
  \bibinfo{booktitle}{\emph{{SIGMOD} '22: International Conference on
  Management of Data, Philadelphia, PA, USA, June 12 - 17, 2022}},
  \bibfield{editor}{\bibinfo{person}{Zachary Ives}, \bibinfo{person}{Angela
  Bonifati}, {and} \bibinfo{person}{Amr~El Abbadi}} (Eds.).
  \bibinfo{publisher}{{ACM}}, \bibinfo{pages}{2246--2258}.
\newblock


\bibitem[\protect\citeauthoryear{Elmasri and Navathe}{Elmasri and
  Navathe}{2015}]%
        {Elmasri_fundamentals_2015}
\bibfield{author}{\bibinfo{person}{Ramez Elmasri} {and}
  \bibinfo{person}{Shamkant~B. Navathe}.} \bibinfo{year}{2015}\natexlab{}.
\newblock \bibinfo{booktitle}{\emph{Fundamentals of Database Systems (7th
  edition)} (\bibinfo{edition}{7th} ed.)}.
\newblock \bibinfo{publisher}{Pearson}.
\newblock
\showISBNx{0133970779}


\bibitem[\protect\citeauthoryear{Facebook}{Facebook}{2018}]%
        {GraphQL2018}
\bibfield{author}{\bibinfo{person}{Facebook}.} \bibinfo{year}{2018}\natexlab{}.
\newblock \bibinfo{title}{{GraphQL}}.
\newblock
\newblock
\newblock
\shownote{https://spec.graphql.org/June2018/.}


\bibitem[\protect\citeauthoryear{Fowler}{Fowler}{2003}]%
        {Fowler2003}
\bibfield{author}{\bibinfo{person}{Martin Fowler}.}
  \bibinfo{year}{2003}\natexlab{}.
\newblock \bibinfo{booktitle}{\emph{UML Distilled: A Brief Guide to the
  Standard Object Modeling Language} (\bibinfo{edition}{3} ed.)}.
\newblock \bibinfo{publisher}{Addison-Wesley Longman Publishing Co., Inc.},
  \bibinfo{address}{USA}.
\newblock
\showISBNx{0321193687}


\bibitem[\protect\citeauthoryear{Francis, Green, Guagliardo, Libkin, Lindaaker,
  Marsault, Plantikow, Rydberg, Selmer, and Taylor}{Francis
  et~al\mbox{.}}{2018}]%
        {cypher}
\bibfield{author}{\bibinfo{person}{Nadime Francis}, \bibinfo{person}{Alastair
  Green}, \bibinfo{person}{Paolo Guagliardo}, \bibinfo{person}{Leonid Libkin},
  \bibinfo{person}{Tobias Lindaaker}, \bibinfo{person}{Victor Marsault},
  \bibinfo{person}{Stefan Plantikow}, \bibinfo{person}{Mats Rydberg},
  \bibinfo{person}{Petra Selmer}, {and} \bibinfo{person}{Andr{\'e}s Taylor}.}
  \bibinfo{year}{2018}\natexlab{}.
\newblock \showarticletitle{Cypher: An Evolving Query Language for Property
  Graphs}. In \bibinfo{booktitle}{\emph{{SIGMOD} Conference}}.
  \bibinfo{publisher}{{ACM}}, \bibinfo{pages}{1433--1445}.
\newblock


\bibitem[\protect\citeauthoryear{Gosnell and Broecheler}{Gosnell and
  Broecheler}{2022}]%
        {PractGraph}
\bibfield{author}{\bibinfo{person}{D.K. Gosnell} {and} \bibinfo{person}{M.
  Broecheler}.} \bibinfo{year}{2022}\natexlab{}.
\newblock \bibinfo{title}{The Practitioner’s Guide to Graph Data}.
\newblock \bibinfo{howpublished}{\url{https://gra.fo/faq/} (visited: 2022-11)}.
\newblock


\bibitem[\protect\citeauthoryear{Gra.fo}{Gra.fo}{2022}]%
        {Grafo}
\bibfield{author}{\bibinfo{person}{Gra.fo}.} \bibinfo{year}{2022}\natexlab{}.
\newblock \bibinfo{title}{Gra.fo}.
\newblock \bibinfo{howpublished}{\url{https://gra.fo/faq/} (visited: 2022-11)}.
\newblock


\bibitem[\protect\citeauthoryear{GraphStudioTM}{GraphStudioTM}{2022}]%
        {TigGraph}
\bibfield{author}{\bibinfo{person}{TigerGraph GraphStudioTM}.}
  \bibinfo{year}{2022}\natexlab{}.
\newblock \bibinfo{title}{TigerGraph GraphStudioTM}.
\newblock
  \bibinfo{howpublished}{\href{https://docs.tigergraph.com/gui/current/graphstudio/overview}{https://docs.tigergraph.com/gui/current/graphstudio/\\overview}
  (visited: 2022-11)}.
\newblock


\bibitem[\protect\citeauthoryear{Group}{Group}{2020}]%
        {LDBC:OAEP:OAEP-2023-04}
\bibfield{author}{\bibinfo{person}{LDBC Property Graph Schema~Working Group}.}
  \bibinfo{year}{2020}\natexlab{}.
\newblock \bibinfo{booktitle}{\emph{LDBC Property Graph Schema contributions to
  WG3}}.
\newblock \bibinfo{type}{Open Access to External Paper} OAEP-2023-04.
  \bibinfo{institution}{Linked Data Benchmark Council (LDBC)}.
\newblock
\urldef\tempurl%
\url{https://doi.org/10.54285/ldbc.OFJF3566}
\showDOI{\tempurl}
\newblock
\shownote{Edited and presented by Jan Hidders, George Fletcher and Bei Li.}


\bibitem[\protect\citeauthoryear{Group, Furniss, and Green}{Group
  et~al\mbox{.}}{2018}]%
        {LDBC:OAEP:OAEP-2023-01}
\bibfield{author}{\bibinfo{person}{Neo4j SQL~Working Group},
  \bibinfo{person}{Peter Furniss}, {and} \bibinfo{person}{Alastair Green}.}
  \bibinfo{year}{2018}\natexlab{}.
\newblock \bibinfo{booktitle}{\emph{SQL/PGQ data model and graph schema}}.
\newblock \bibinfo{type}{Open Access to External Paper} OAEP-2023-01.
  \bibinfo{institution}{Linked Data Benchmark Council (LDBC)}.
\newblock
\urldef\tempurl%
\url{https://doi.org/10.54285/ldbc.QZSK3559}
\showDOI{\tempurl}


\bibitem[\protect\citeauthoryear{Groz, Lemay, Staworko, and Wieczorek}{Groz
  et~al\mbox{.}}{2022}]%
        {GrozLSW22}
\bibfield{author}{\bibinfo{person}{Beno{\^{\i}}t Groz},
  \bibinfo{person}{Aur{\'{e}}lien Lemay}, \bibinfo{person}{Slawek Staworko},
  {and} \bibinfo{person}{Piotr Wieczorek}.} \bibinfo{year}{2022}\natexlab{}.
\newblock \showarticletitle{Inference of Shape Graphs for Graph Databases}. In
  \bibinfo{booktitle}{\emph{25th International Conference on Database Theory,
  {ICDT} 2022, March 29 to April 1, 2022, Edinburgh, {UK} (Virtual
  Conference)}} \emph{(\bibinfo{series}{LIPIcs}, Vol.~\bibinfo{volume}{220})},
  \bibfield{editor}{\bibinfo{person}{Dan Olteanu} {and} \bibinfo{person}{Nils
  Vortmeier}} (Eds.). \bibinfo{publisher}{Schloss Dagstuhl - Leibniz-Zentrum
  f{\"{u}}r Informatik}, \bibinfo{pages}{14:1--14:20}.
\newblock
\urldef\tempurl%
\url{https://doi.org/10.4230/LIPIcs.ICDT.2022.14}
\showDOI{\tempurl}


\bibitem[\protect\citeauthoryear{Halpin}{Halpin}{2015}]%
        {halpin2015}
\bibfield{author}{\bibinfo{person}{Terry Halpin}.}
  \bibinfo{year}{2015}\natexlab{}.
\newblock \bibinfo{booktitle}{\emph{Object-{Role} {Modeling} {Fundamentals}:
  {A} {Practical} {Guide} to {Data} {Modeling} with {ORM}}}.
\newblock \bibinfo{publisher}{Technics Publications}.
\newblock
\showISBNx{978-1-63462-074-1}


\bibitem[\protect\citeauthoryear{Hartig and Hidders}{Hartig and
  Hidders}{2019}]%
        {GraphQLforPG}
\bibfield{author}{\bibinfo{person}{Olaf Hartig} {and} \bibinfo{person}{Jan
  Hidders}.} \bibinfo{year}{2019}\natexlab{}.
\newblock \showarticletitle{Defining Schemas for Property Graphs by Using the
  GraphQL Schema Definition Language}. In \bibinfo{booktitle}{\emph{Proceedings
  of the 2nd Joint International Workshop on Graph Data Management Experiences
  \& Systems (GRADES) and Network Data Analytics (NDA)}} (Amsterdam,
  Netherlands) \emph{(\bibinfo{series}{GRADES-NDA'19})}.
  \bibinfo{publisher}{Association for Computing Machinery},
  \bibinfo{address}{New York, NY, USA}, Article \bibinfo{articleno}{6},
  \bibinfo{numpages}{11}~pages.
\newblock
\showISBNx{9781450367899}
\urldef\tempurl%
\url{https://doi.org/10.1145/3327964.3328495}
\showDOI{\tempurl}


\bibitem[\protect\citeauthoryear{Hartig and P{\'e}rez}{Hartig and
  P{\'e}rez}{2018}]%
        {Hartig2018semantics}
\bibfield{author}{\bibinfo{person}{Olaf Hartig} {and} \bibinfo{person}{Jorge
  P{\'e}rez}.} \bibinfo{year}{2018}\natexlab{}.
\newblock \showarticletitle{Semantics and complexity of GraphQL}. In
  \bibinfo{booktitle}{\emph{Proceedings of the 2018 World Wide Web
  Conference}}. \bibinfo{pages}{1155--1164}.
\newblock


\bibitem[\protect\citeauthoryear{Hitzler, Rudolph, Kr{\"{o}}tzsch,
  Patel-Schneider, and Parsia}{Hitzler et~al\mbox{.}}{2012}]%
        {Hitzler2012OWO}
\bibfield{author}{\bibinfo{person}{Pascal Hitzler}, \bibinfo{person}{Sebastian
  Rudolph}, \bibinfo{person}{Markus Kr{\"{o}}tzsch}, \bibinfo{person}{Peter
  Patel-Schneider}, {and} \bibinfo{person}{Bijan Parsia}.}
  \bibinfo{year}{2012}\natexlab{}.
\newblock \bibinfo{booktitle}{\emph{{OWL} 2 Web Ontology Language Primer
  (Second Edition)}}.
\newblock \bibinfo{type}{{W3C} Recommendation}. \bibinfo{institution}{W3C}.
\newblock
\newblock
\shownote{\url{https://www.w3.org/TR/2012/REC-owl2-primer-20121211/}.}


\bibitem[\protect\citeauthoryear{ISO/IEC 19757-2:2008}{ISO/IEC
  19757-2:2008}{2008}]%
        {ISO19757-2}
ISO/IEC 19757-2:2008 \bibinfo{year}{2008}\natexlab{}.
\newblock \bibinfo{booktitle}{\emph{{Information technology — Document Schema
  Definition Language (DSDL) — Part 2: Regular-grammar-based validation —
  RELAX NG}}}.
\newblock \bibinfo{type}{Standard}. \bibinfo{institution}{International
  Organization for Standardization}, \bibinfo{address}{Geneva, CH}.
\newblock


\bibitem[\protect\citeauthoryear{JanusGraph}{JanusGraph}{2022}]%
        {JanusDB}
\bibfield{author}{\bibinfo{person}{JanusGraph}.}
  \bibinfo{year}{2022}\natexlab{}.
\newblock \bibinfo{title}{JanusGraph}.
\newblock \bibinfo{howpublished}{\url{https://janusgraph.org/} (visited:
  2022-11)}.
\newblock


\bibitem[\protect\citeauthoryear{Knublauch and Kontokostas}{Knublauch and
  Kontokostas}{2017}]%
        {Knublauch2017SCL}
\bibfield{author}{\bibinfo{person}{Holger Knublauch} {and}
  \bibinfo{person}{Dimitris Kontokostas}.} \bibinfo{year}{2017}\natexlab{}.
\newblock \bibinfo{booktitle}{\emph{Shapes Constraint Language ({SHACL})}}.
\newblock \bibinfo{type}{{W3C} Recommendation}. \bibinfo{institution}{W3C}.
\newblock
\newblock
\shownote{\url{https://www.w3.org/TR/2017/REC-shacl-20170720/}.}


\bibitem[\protect\citeauthoryear{Martens, Neven, Niewerth, and
  Schwentick}{Martens et~al\mbox{.}}{2012}]%
        {DBLP:journals/pvldb/MartensNNS12}
\bibfield{author}{\bibinfo{person}{Wim Martens}, \bibinfo{person}{Frank Neven},
  \bibinfo{person}{Matthias Niewerth}, {and} \bibinfo{person}{Thomas
  Schwentick}.} \bibinfo{year}{2012}\natexlab{}.
\newblock \showarticletitle{Developing and Analyzing XSDs through BonXai}.
\newblock \bibinfo{journal}{\emph{Proc. {VLDB} Endow.}} \bibinfo{volume}{5},
  \bibinfo{number}{12} (\bibinfo{year}{2012}), \bibinfo{pages}{1994--1997}.
\newblock
\urldef\tempurl%
\url{https://doi.org/10.14778/2367502.2367556}
\showDOI{\tempurl}


\bibitem[\protect\citeauthoryear{Martens, Neven, Niewerth, and
  Schwentick}{Martens et~al\mbox{.}}{2017}]%
        {DBLP:journals/tods/MartensNNS17}
\bibfield{author}{\bibinfo{person}{Wim Martens}, \bibinfo{person}{Frank Neven},
  \bibinfo{person}{Matthias Niewerth}, {and} \bibinfo{person}{Thomas
  Schwentick}.} \bibinfo{year}{2017}\natexlab{}.
\newblock \showarticletitle{BonXai: Combining the Simplicity of {DTD} with the
  Expressiveness of {XML} Schema}.
\newblock \bibinfo{journal}{\emph{{ACM} Trans. Database Syst.}}
  \bibinfo{volume}{42}, \bibinfo{number}{3} (\bibinfo{year}{2017}),
  \bibinfo{pages}{15:1--15:42}.
\newblock


\bibitem[\protect\citeauthoryear{Needham and Hodler}{Needham and
  Hodler}{2019}]%
        {NeHo19}
\bibfield{author}{\bibinfo{person}{Mark Needham} {and} \bibinfo{person}{Amy~E.
  Hodler}.} \bibinfo{year}{2019}\natexlab{}.
\newblock \bibinfo{booktitle}{\emph{Graph {Algorithms}}}.
\newblock \bibinfo{publisher}{O'Relly Media}.
\newblock


\bibitem[\protect\citeauthoryear{Neo4j}{Neo4j}{2016}]%
        {Neo4j16}
Neo4j \bibinfo{year}{2016}\natexlab{}.
\newblock \bibinfo{booktitle}{\emph{The Definitive Guide to Graph Databases for
  the {RDBMS} Developer}}.
\newblock Neo4j.
\newblock


\bibitem[\protect\citeauthoryear{Neo4j}{Neo4j}{2019}]%
        {morpheus}
\bibfield{author}{\bibinfo{person}{Neo4j}.} \bibinfo{year}{2019}\natexlab{}.
\newblock \bibinfo{title}{Graph {DDL} (Data Definition Language)}.
\newblock
  \bibinfo{howpublished}{\href{https://github.com/opencypher/morpheus/blob/master/documentation/asciidoc/backend-sql-graphddl.adoc}{https://github.com/opencypher/morpheus/blob/master/\\documentation/asciidoc/backend-sql-graphddl.adoc}
  (visited: 2023-04)}.
\newblock


\bibitem[\protect\citeauthoryear{Neo4j}{Neo4j}{2022a}]%
        {Neo4jDB}
\bibfield{author}{\bibinfo{person}{Neo4j}.} \bibinfo{year}{2022}\natexlab{a}.
\newblock \bibinfo{title}{Neo4j}.
\newblock \bibinfo{howpublished}{\url{https://neo4j.com/} (visited: 2022-11)}.
\newblock


\bibitem[\protect\citeauthoryear{Neo4j}{Neo4j}{2022b}]%
        {NeoB}
\bibfield{author}{\bibinfo{person}{Neo4j}.} \bibinfo{year}{2022}\natexlab{b}.
\newblock \bibinfo{title}{Neo4j Browser}.
\newblock
  \bibinfo{howpublished}{\url{https://neo4j.com/product/developer-tools/\#browser}
  (visited: 2022-11)}.
\newblock


\bibitem[\protect\citeauthoryear{Notebooks}{Notebooks}{2022}]%
        {GraphNB}
\bibfield{author}{\bibinfo{person}{Graph Notebooks}.}
  \bibinfo{year}{2022}\natexlab{}.
\newblock \bibinfo{title}{Graph Notebooks}.
\newblock \bibinfo{howpublished}{\url{https://github.com/aws/graph-notebook}
  (visited: 2022-11)}.
\newblock


\bibitem[\protect\citeauthoryear{Oracle}{Oracle}{2022}]%
        {OracleSGDB}
\bibfield{author}{\bibinfo{person}{Oracle}.} \bibinfo{year}{2022}\natexlab{}.
\newblock \bibinfo{title}{Oracle Spatial and Graph}.
\newblock
  \bibinfo{howpublished}{\url{https://www.oracle.com/database/technologies/spatialandgraph.html}
  (visited: 2022-11)}.
\newblock


\bibitem[\protect\citeauthoryear{OrientDB}{OrientDB}{2022}]%
        {OrientDB}
\bibfield{author}{\bibinfo{person}{OrientDB}.} \bibinfo{year}{2022}\natexlab{}.
\newblock \bibinfo{title}{OrientDB}.
\newblock \bibinfo{howpublished}{\url{https://orientdb.org/} (visited:
  2022-11)}.
\newblock


\bibitem[\protect\citeauthoryear{Peterson, Gao, Biron, Sperberg-McQueen,
  Malhotra, and Thompson}{Peterson et~al\mbox{.}}{2012}]%
        {Peterson2012WXS}
\bibfield{author}{\bibinfo{person}{David Peterson}, \bibinfo{person}{Sandy
  Gao}, \bibinfo{person}{Paul~V. Biron}, \bibinfo{person}{Michael
  Sperberg-McQueen}, \bibinfo{person}{Ashok Malhotra}, {and}
  \bibinfo{person}{Henry Thompson}.} \bibinfo{year}{2012}\natexlab{}.
\newblock \bibinfo{booktitle}{\emph{{W3C} XML Schema Definition Language
  ({XSD}) {1.1} Part 2: Datatypes}}.
\newblock \bibinfo{type}{{W3C} Recommendation}. \bibinfo{institution}{W3C}.
\newblock
\newblock
\shownote{\url{https://www.w3.org/TR/2012/REC-xmlschema11-2-20120405/}.}


\bibitem[\protect\citeauthoryear{Robinson, Webber, and Eifrem}{Robinson
  et~al\mbox{.}}{2015}]%
        {RoWeEi15}
\bibfield{author}{\bibinfo{person}{Ian Robinson}, \bibinfo{person}{Jim Webber},
  {and} \bibinfo{person}{Emil Eifrem}.} \bibinfo{year}{2015}\natexlab{}.
\newblock \bibinfo{booktitle}{\emph{Graph Databases}}.
\newblock \bibinfo{publisher}{O'Reilly Media}.
\newblock


\bibitem[\protect\citeauthoryear{Rydberg}{Rydberg}{2016}]%
        {LDBC:OAEP:OAEP-2023-03}
\bibfield{author}{\bibinfo{person}{Mats Rydberg}.}
  \bibinfo{year}{2016}\natexlab{}.
\newblock \bibinfo{booktitle}{\emph{Cypher schema constraints proposal}}.
\newblock \bibinfo{type}{Open Access to External Paper} OAEP-2023-03.
  \bibinfo{institution}{Linked Data Benchmark Council (LDBC)}.
\newblock
\urldef\tempurl%
\url{https://doi.org/10.54285/ldbc.KKHM1756}
\showDOI{\tempurl}


\bibitem[\protect\citeauthoryear{Sahu, Mhedhbi, Salihoglu, Lin, and
  {\"{O}}zsu}{Sahu et~al\mbox{.}}{2020a}]%
        {SahuMSLO20}
\bibfield{author}{\bibinfo{person}{Siddhartha Sahu}, \bibinfo{person}{Amine
  Mhedhbi}, \bibinfo{person}{Semih Salihoglu}, \bibinfo{person}{Jimmy Lin},
  {and} \bibinfo{person}{M.~Tamer {\"{O}}zsu}.}
  \bibinfo{year}{2020}\natexlab{a}.
\newblock \showarticletitle{The ubiquity of large graphs and surprising
  challenges of graph processing: extended survey}.
\newblock \bibinfo{journal}{\emph{{VLDB} J.}} \bibinfo{volume}{29},
  \bibinfo{number}{2-3} (\bibinfo{year}{2020}), \bibinfo{pages}{595--618}.
\newblock


\bibitem[\protect\citeauthoryear{Sahu, Mhedhbi, Salihoglu, Lin, and
  {\"O}zsu}{Sahu et~al\mbox{.}}{2020b}]%
        {MSSJO20}
\bibfield{author}{\bibinfo{person}{Siddhartha Sahu}, \bibinfo{person}{Amine
  Mhedhbi}, \bibinfo{person}{Semih Salihoglu}, \bibinfo{person}{Jimmy Lin},
  {and} \bibinfo{person}{M.~Tamer {\"O}zsu}.} \bibinfo{year}{2020}\natexlab{b}.
\newblock \showarticletitle{The ubiquity of large graphs and surprising
  challenges of graph processing: extended survey}.
\newblock \bibinfo{journal}{\emph{The VLDB Journal}} \bibinfo{volume}{29},
  \bibinfo{number}{2} (\bibinfo{year}{2020}), \bibinfo{pages}{595--618}.
\newblock
\showISBNx{0949-877X}
\urldef\tempurl%
\url{https://doi.org/10.1007/s00778-019-00548-x}
\showDOI{\tempurl}


\bibitem[\protect\citeauthoryear{Sakr, Bonifati, Voigt, Iosup, Ammar, Angles,
  Aref, Arenas, Besta, Boncz, Daudjee, Valle, Dumbrava, Hartig, Haslhofer,
  Hegeman, Hidders, Hose, Iamnitchi, Kalavri, Kapp, Martens, {\"{O}}zsu,
  Peukert, Plantikow, Ragab, Ripeanu, Salihoglu, Schulz, Selmer, Sequeda,
  Shinavier, Sz{\'{a}}rnyas, Tommasini, Tumeo, Uta, Varbanescu, Wu, Yakovets,
  Yan, and Yoneki}{Sakr et~al\mbox{.}}{2021}]%
        {SakrBVIAAAABBDV21}
\bibfield{author}{\bibinfo{person}{Sherif Sakr}, \bibinfo{person}{Angela
  Bonifati}, \bibinfo{person}{Hannes Voigt}, \bibinfo{person}{Alexandru Iosup},
  \bibinfo{person}{Khaled Ammar}, \bibinfo{person}{Renzo Angles},
  \bibinfo{person}{Walid~G. Aref}, \bibinfo{person}{Marcelo Arenas},
  \bibinfo{person}{Maciej Besta}, \bibinfo{person}{Peter~A. Boncz},
  \bibinfo{person}{Khuzaima Daudjee}, \bibinfo{person}{Emanuele~Della Valle},
  \bibinfo{person}{Stefania Dumbrava}, \bibinfo{person}{Olaf Hartig},
  \bibinfo{person}{Bernhard Haslhofer}, \bibinfo{person}{Tim Hegeman},
  \bibinfo{person}{Jan Hidders}, \bibinfo{person}{Katja Hose},
  \bibinfo{person}{Adriana Iamnitchi}, \bibinfo{person}{Vasiliki Kalavri},
  \bibinfo{person}{Hugo Kapp}, \bibinfo{person}{Wim Martens},
  \bibinfo{person}{M.~Tamer {\"{O}}zsu}, \bibinfo{person}{Eric Peukert},
  \bibinfo{person}{Stefan Plantikow}, \bibinfo{person}{Mohamed Ragab},
  \bibinfo{person}{Matei Ripeanu}, \bibinfo{person}{Semih Salihoglu},
  \bibinfo{person}{Christian Schulz}, \bibinfo{person}{Petra Selmer},
  \bibinfo{person}{Juan~F. Sequeda}, \bibinfo{person}{Joshua Shinavier},
  \bibinfo{person}{G{\'{a}}bor Sz{\'{a}}rnyas}, \bibinfo{person}{Riccardo
  Tommasini}, \bibinfo{person}{Antonino Tumeo}, \bibinfo{person}{Alexandru
  Uta}, \bibinfo{person}{Ana~Lucia Varbanescu}, \bibinfo{person}{Hsiang{-}Yun
  Wu}, \bibinfo{person}{Nikolay Yakovets}, \bibinfo{person}{Da Yan}, {and}
  \bibinfo{person}{Eiko Yoneki}.} \bibinfo{year}{2021}\natexlab{}.
\newblock \showarticletitle{The future is big graphs: a community view on graph
  processing systems}.
\newblock \bibinfo{journal}{\emph{Commun. {ACM}}} \bibinfo{volume}{64},
  \bibinfo{number}{9} (\bibinfo{year}{2021}), \bibinfo{pages}{62--71}.
\newblock


\bibitem[\protect\citeauthoryear{Sawyer}{Sawyer}{2022}]%
        {TSDB}
\bibfield{author}{\bibinfo{person}{Tom Sawyer}.}
  \bibinfo{year}{2022}\natexlab{}.
\newblock \bibinfo{title}{Graph Database Browser}.
\newblock
  \bibinfo{howpublished}{\url{https://www.tomsawyer.com/graph-database-browser}
  (visited: 2022-11)}.
\newblock


\bibitem[\protect\citeauthoryear{Seifer, L\"{a}mmel, and Staab}{Seifer
  et~al\mbox{.}}{2021}]%
        {Seifer21}
\bibfield{author}{\bibinfo{person}{Philipp Seifer}, \bibinfo{person}{Ralf
  L\"{a}mmel}, {and} \bibinfo{person}{Steffen Staab}.}
  \bibinfo{year}{2021}\natexlab{}.
\newblock \showarticletitle{ProGS: Property Graph Shapes Language}. In
  \bibinfo{booktitle}{\emph{The Semantic Web – ISWC 2021: 20th International
  Semantic Web Conference, ISWC 2021, Virtual Event, October 24–28, 2021,
  Proceedings}}. \bibinfo{publisher}{Springer-Verlag},
  \bibinfo{address}{Berlin, Heidelberg}, \bibinfo{pages}{392–409}.
\newblock
\showISBNx{978-3-0r30-88360-7}
\urldef\tempurl%
\url{https://doi.org/10.1007/978-3-030-88361-4_23}
\showDOI{\tempurl}


\bibitem[\protect\citeauthoryear{Sperberg-McQueen, Thompson, Beech, Maloney,
  Mendelsohn, and Gao}{Sperberg-McQueen et~al\mbox{.}}{2012}]%
        {SperbergMcQueen2012WXS}
\bibfield{author}{\bibinfo{person}{Michael Sperberg-McQueen},
  \bibinfo{person}{Henry Thompson}, \bibinfo{person}{David Beech},
  \bibinfo{person}{Murray Maloney}, \bibinfo{person}{Noah Mendelsohn}, {and}
  \bibinfo{person}{Sandy Gao}.} \bibinfo{year}{2012}\natexlab{}.
\newblock \bibinfo{booktitle}{\emph{{W3C} XML Schema Definition Language
  ({XSD}) {1.1} Part 1: Structures}}.
\newblock \bibinfo{type}{{W3C} Recommendation}. \bibinfo{institution}{W3C}.
\newblock
\newblock
\shownote{\url{https://www.w3.org/TR/2012/REC-xmlschema11-1-20120405/}.}


\bibitem[\protect\citeauthoryear{Staworko, Boneva, Gayo, Hym, Prud'Hommeaux,
  and Solbrig}{Staworko et~al\mbox{.}}{2015}]%
        {Staworko2015complexity}
\bibfield{author}{\bibinfo{person}{Slawomir Staworko}, \bibinfo{person}{Iovka
  Boneva}, \bibinfo{person}{Jos{\'e} Emilio~Labra Gayo},
  \bibinfo{person}{Samuel Hym}, \bibinfo{person}{Eric~Gordon Prud'Hommeaux},
  {and} \bibinfo{person}{Harold Solbrig}.} \bibinfo{year}{2015}\natexlab{}.
\newblock \showarticletitle{Complexity and Expressiveness of ShEx for RDF}. In
  \bibinfo{booktitle}{\emph{18th International Conference on Database Theory
  (ICDT 2015)}}.
\newblock


\bibitem[\protect\citeauthoryear{Team}{Team}{2019}]%
        {LDBC:OAEP:OAEP-2023-02}
\bibfield{author}{\bibinfo{person}{Neo4j Query Languages Standards \&~Research
  Team}.} \bibinfo{year}{2019}\natexlab{}.
\newblock \bibinfo{booktitle}{\emph{Introduction to GQL Schema design}}.
\newblock \bibinfo{type}{Open Access to External Paper} OAEP-2023-02.
  \bibinfo{institution}{Linked Data Benchmark Council (LDBC)}.
\newblock
\urldef\tempurl%
\url{https://doi.org/10.54285/ldbc.EPWQ6741}
\showDOI{\tempurl}
\newblock
\shownote{Edited by Alastair Green and Hannes Voigt.}


\bibitem[\protect\citeauthoryear{Technologies}{Technologies}{2022}]%
        {SparkseeDB}
\bibfield{author}{\bibinfo{person}{Sparsity Technologies}.}
  \bibinfo{year}{2022}\natexlab{}.
\newblock \bibinfo{title}{Sparksee}.
\newblock
  \bibinfo{howpublished}{\url{https://sparsity-technologies.com/\#sparksee}
  (visited: 2022-11)}.
\newblock


\bibitem[\protect\citeauthoryear{Thalheim}{Thalheim}{2018}]%
        {Thalheim_extended_2018}
\bibfield{author}{\bibinfo{person}{Bernhard Thalheim}.}
  \bibinfo{year}{2018}\natexlab{}.
\newblock \showarticletitle{Extended {Entity}-{Relationship} {Model}}.
\newblock In \bibinfo{booktitle}{\emph{Encyclopedia of {Database} {Systems},
  {Second} {Edition}}}, \bibfield{editor}{\bibinfo{person}{Ling Liu} {and}
  \bibinfo{person}{M.~Tamer {\"O}zsu}} (Eds.). \bibinfo{publisher}{Springer}.
\newblock
\urldef\tempurl%
\url{https://doi.org/10.1007/978-1-4614-8265-9_157}
\showDOI{\tempurl}


\bibitem[\protect\citeauthoryear{TigerGraph}{TigerGraph}{2022}]%
        {TigerGraphDB}
\bibfield{author}{\bibinfo{person}{TigerGraph}.}
  \bibinfo{year}{2022}\natexlab{}.
\newblock \bibinfo{title}{TigerGraph}.
\newblock \bibinfo{howpublished}{\url{https://www.tigergraph.com/} (visited:
  2022-11)}.
\newblock


\bibitem[\protect\citeauthoryear{Vaticle}{Vaticle}{2022}]%
        {TypeDB}
\bibfield{author}{\bibinfo{person}{Vaticle}.} \bibinfo{year}{2022}\natexlab{}.
\newblock \bibinfo{title}{TypeDB}.
\newblock \bibinfo{howpublished}{\url{https://vaticle.com/} (visited:
  2022-11)}.
\newblock


\bibitem[\protect\citeauthoryear{Wileński and Tomaszuk}{Wileński and
  Tomaszuk}{2022}]%
        {dw2022}
\bibfield{author}{\bibinfo{person}{Damian Wileński} {and}
  \bibinfo{person}{Dominik Tomaszuk}.} \bibinfo{year}{2022}\natexlab{}.
\newblock \showarticletitle{damianw27/pgs-grammar-check: Version 1.0.0}.
\newblock  (\bibinfo{date}{Nov} \bibinfo{year}{2022}).
\newblock
\urldef\tempurl%
\url{https://doi.org/10.5281/zenodo.7344227}
\showDOI{\tempurl}
\newblock
\shownote{Available at \url{https://damianw27.github.io/pgs-grammar-check/}.}


\bibitem[\protect\citeauthoryear{Wright, Andrews, Hutton, and Dennis}{Wright
  et~al\mbox{.}}{2020}]%
        {Wright2020JS}
\bibfield{author}{\bibinfo{person}{Austin Wright}, \bibinfo{person}{Henry
  Andrews}, \bibinfo{person}{Ben Hutton}, {and} \bibinfo{person}{Greg Dennis}.}
  \bibinfo{year}{2020}\natexlab{}.
\newblock \bibinfo{booktitle}{\emph{{JSON} Schema: A Media Type for Describing
  JSON Documents}}.
\newblock \bibinfo{type}{Draft}. \bibinfo{institution}{Internet Engineering
  Task Force}.
\newblock


\bibitem[\protect\citeauthoryear{Wu, Yi, Yu, Liu, and Wang}{Wu
  et~al\mbox{.}}{2022}]%
        {NebulaDB}
\bibfield{author}{\bibinfo{person}{Min Wu}, \bibinfo{person}{Xinglu Yi},
  \bibinfo{person}{Hui Yu}, \bibinfo{person}{Yu Liu}, {and}
  \bibinfo{person}{Yujue Wang}.} \bibinfo{year}{2022}\natexlab{}.
\newblock \showarticletitle{Nebula Graph: An open source distributed graph
  database}.
\newblock \bibinfo{journal}{\emph{CoRR}}  \bibinfo{volume}{abs/2206.07278}
  (\bibinfo{year}{2022}).
\newblock


\bibitem[\protect\citeauthoryear{Yergeau, Sperberg-McQueen, Bray, Paoli, and
  Maler}{Yergeau et~al\mbox{.}}{2008}]%
        {Yergeau2008EML}
\bibfield{author}{\bibinfo{person}{Fran\c{c}ois Yergeau},
  \bibinfo{person}{Michael Sperberg-McQueen}, \bibinfo{person}{Tim Bray},
  \bibinfo{person}{Jean Paoli}, {and} \bibinfo{person}{Eve Maler}.}
  \bibinfo{year}{2008}\natexlab{}.
\newblock \bibinfo{booktitle}{\emph{Extensible Markup Language ({XML}) 1.0
  (Fifth Edition)}}.
\newblock \bibinfo{type}{{W3C} Recommendation}. \bibinfo{institution}{W3C}.
\newblock
\newblock
\shownote{https://www.w3.org/TR/2008/REC-xml-20081126/.}


\end{thebibliography}

%\clearpage 
%\appendix
%\input{7-appendix.tex}

\end{document}